\documentclass[10pt,showpacs]{revtex4}
\usepackage{amsmath,amssymb,amsfonts,epsfig,psfrag,graphicx}
\begin{document}

\title{Thermodynamics of the quantum spin-$S$ $XXZ$ chain}

\date{\today}

\author{Onofre Rojas and S.M. de Souza}

\affiliation{Departamento de Ci\^encias Exatas, Universidade Federal de
Lavras,  PO-Box 37, CEP 37200-000, Lavras-MG,  Brazil}

\author{E.V. Corr\^ea Silva}
\affiliation{Departamento de Matem\'atica e Computa\c c\~ao,
Faculdade de Tecnologia,
Universidade do Estado do Rio de Janeiro.
Estrada Resende-Riachuelo s/n$^{\textit o}$,
Morada da Colina, CEP 27523-000,  Resende-RJ, Brazil}

\author{M.T. Thomaz}
\affiliation{Instituto de F\'{\i}sica,
Universidade Federal Fluminense, Av. Gal. Milton Tavares de
Souza s/n$^{\textit o}$, CEP 24210-340, Niter\'oi - RJ, Brazil}

\begin{abstract}
The thermodynamics  of the spin-$S$ anisotropic quantum $XXZ$ chain
with  arbitrary  value of $S$ and unitary norm,
 in the high-temperature regime, is
reported. The  single-ion anisotropy term
and the interaction with an external magnetic field in 
the $z$-direction are taken
into account. We obtain, for arbitrary value of $S$, the 
$\beta$-expansion of the Helmholtz free energy of
the model  up to order $\beta^6$ and show that it actually 
depends on $\frac{1}{S(S+1)}$.  Its classical limit 
 is obtained by simply taking $S\rightarrow \infty$.
 At $h=0$ and $D=0$, our high temperature expansion of the
 classical model coincides with Joyce's exact solution\cite{joyce_prl}.
 We study, in the high temperature region, some thermodynamic quantities
such as the specific heat and the magnetic susceptibility as
functions of spin and verify
for which values of $S$ those thermodynamic functions behave classically. 
Their finite temperature behavior is inferred from interpolation of 
their high- and low-temperature behavior,
and shown to be in good agreement with numerical results. 
The finite temperature behavior is shown for higher values of spin.
\end{abstract}


\maketitle

\section{Introduction}

One-dimensional and quasi-one-dimensional quantum chain models 
have been intensively investigated 
for a twofold reason. On one hand, for the
algebraic properties of spin-$1/2$ chain models 
($XXZ$ and generalized ladder models)
which, being  integrable   models, can be solved exactly 
by the powerful Bethe {\it ansatz} technique (see Refs. \cite{juttner}
and \cite{angela}, respectively). On the other hand, for
the existence of quasi one-dimensional 
magnetic systems that exhibit nearly ideal one-dimensional behavior 
for some interval of temperature. The thermodynamic properties of those 
magnetic systems have been described by
spin-$S$  $XXZ$  models for various values of 
spin, mixed-spin  models, ladder models,  etc. 
In particular, there are materials
well described by higher spin values of the $XXZ$ models
such as ${\rm CsVCl}_3$ and ${\rm CsVBr}_3$ 
($S=3/2$)\cite{kadowaki,itoh1,itoh2},  
$({\rm C}_{10} {\rm H}_8{\rm N}_2){\rm MnCl}_3$
($S=2$)\cite{granroth} and $({\rm CD}_3)_4{\rm NMnCl}_3$  
$(S=5/2)$\cite{birgeneau,hutchings}. Motivated by the discovery of these 
materials, the magnetic and  thermodynamic  properties 
of the ferro- and antiferromagnetic isotropic spin-$S$ $XXZ$ model
for higher values of $S$  
have  been studied numerically
and analytically (e.g., the high temperature
expansions; see Ref. \cite{jpcm03} and references therein).

The classical limit ($S \rightarrow \infty$) of the
$XXZ$ model is exactly soluble and was originally solved  
by Fisher\cite{fisher}, in the isotropic
regime, for a null external magnetic field and no single-ion anisotropy term.
Its anysotropic case has been solved by 
Joyce \cite{joyce_prl}, also in the absence of an external magnetic
field and without a  single-ion  anisotropy term, by writing the Helmholtz
 free energy of the classical $XXZ$ 
model as an integral equation  whose solution is the 
spheroidal wave function. Unfortunately, in the
presence of an external magnetic field and/or the single-ion anisotropy
parameter, the solution of the integral equation 
cannot be reduced to  any known function. 
 
Recently, Fukushima {\it et al.}\cite{fukushima} obtained
the specific heat and the magnetic
susceptibility per site of a ferromagnetic mixed-spin model, with two
kinds of spins, $s$ and $S$, arranged alternatively and
coupled by a Heisenberg-type nearest-neighbor exchange for
arbitrary values of $s$ and $S$ at the isotropic point
($\Delta =1$ in eq.(\ref{hamt1})) and in the absence of external
magnetic fields. They obtained numerical results
through the exact diagonalization method and the analytical high
temperature expansions.  For $s=S$ their model
becomes a one-dimensional spin-$S$
$XXZ$ model for arbitrary value of $S$. Their high temperature
expansion of the specific heat per site, in the absence
 of an external magnetic field, goes up to order
$(\beta J)^{11}$ and their expansion of the magnetic
susceptibility, also for a  vanishing magnetic field,
goes up to $(\beta J)^7$   with a  single-ion anisotropy term in one
of the spins.  In Ref. \cite{jpcm03} we extended  the 
results of Ref. \cite{fukushima}, for spin values up to $S=4$,
including  a free  parameter of anisotropy 
in the $z$-direction,  a single-ion anisotropy term and a
non-zero external magnetic field in the $\beta$-expansions of the
Helmholtz free energies up to order $(J\beta)^6$.

Following Ref. \cite{jpcm03}, we also consider the  Hamiltonian of
the  anisotropic spin-$S$ $XXZ$ with a single-ion anisotropy term  and in the
presence of an external magnetic field, 

\begin{align}\label{hamt1}
{\bf H} = \sum_{i=1}^N J' \left( {\bf S}_{i}, {\bf S}_{i+1}\right)_{\Delta}
       - h' S^{z}_{i} + D' (S^{z}_{i})^2.
\end{align}

\noindent We use the notation:
$( {\bf S}_{i}, {\bf S}_{i+1})_{\Delta} \equiv
S^{x}_{i} S^{x}_{i+1} + S^{y}_{i} S^{y}_{i+1} + \Delta S^{z}_{i}
S^{z}_{i+1}$. Here, $S^x_i$, $S^y_i$ and $S^z_i$ stand for the
spin-$S$ matrices in the {\it i}-th site of the chain and norm 
$\sqrt{S(S+1)}$; $N$ is the  number of sites in the periodic
 chain; $J'$ is the exchange integral; $\Delta$ is the anisotropy
constant in the $z$-direction; $h'$ is the external magnetic field
in the $z$-axis and $D'$ is the single-ion anisotropy parameter.

Even in one spatial dimension, the solution of the quantum chain model
(\ref{hamt1}) is more complex if we consider a finite arbitrary
spin-$S$ (with $S\not= 1/2$) once the $XXZ$ model becomes a
non-integrable model and cannot be solved by the thermodynamic Bethe
{\it ansatz} technique.

Numerical  approaches for higher spin values becomes more involved 
since there are  more degrees of freedom to  be handled.
Certainly,  the  high temperature expansions are easily
 manipulated  by  symbolic computer languages
 and can be used as a  reliable  check  for
 numerical calculations in the high temperature regime.

The aim of the present communication is to extend the results of Ref.
\cite{jpcm03} to arbitrary values of the spin $S$ in hamiltonian
(\ref{hamt1}). In doing so, we are including in the results of Ref.
\cite{fukushima} the effects of anisotropies in the $z$-direction of
the $XXZ$ model in the presence of an external magnetic field. Having
a high temperature expansion for the model, for any $S$, we can derive
its classical limit in this regime of temperature. We want to
re-obtain the classical results of the Heisenberg model by considering
the quantum nature of the spin variable, in opposition to the known
results in the literature where this limit is derived from a chain of
classical spins. From the high temperature expansion 
of the Helmholtz free energy, we can also verify for
which values of $S$ the anisotropic quantum $XXZ$ model, in the
presence of an external magnetic field, is well described by its
classical version, extending the results of Ref. \cite{jpcm03}. 
In the present article, comparison of thermodynamic functions for
different values of $S$ is performed at the same (high) $T$, whereas
in Ref. \cite{jpcm03}, they were plotted as functions of the
``rescaled'' temperature $\tilde{T} = \frac{T}{S(S+1)}$.

This paper is organized as follows: in \S II we use the method
developed in Ref. \cite{chain_m} and an interpolation technique to
obtain the high-temperature series expansion of the Helmholtz free
energy for the quantum spin-$s$ XXZ chain ($s$ with unitary norm) up
to order $\beta^6$. In order to check our analytic $\beta$-expansion,
in the first subsection of \S II our results for the classical limit
($S\rightarrow\infty$) are compared to the well-known
isotropic\cite{fisher} and anisotropic\cite{joyce_prl} classical
Heisenberg chains. In the second subsection of \S II our results for
finite values of $S$ are explored to include the effect of the
anisotropy parameter $\Delta$, the presence of $D^\prime$ and
$h^\prime$ in the Hamiltonian \eqref{hamt1} and the dependence of
thermodynamic functions on $S$. In \S III we use Pad\'e
representatives to enhance our high-temperature expansions for the
spin-$s$ $XXZ$ model, extending them to lower temperatures. We do so
by taking into account the known behavior at $T\sim0$. For the
specific heat we use the method presented in Ref. \cite{bernu},
whereas for the magnetic susceptibility we apply the Dlog-Pad\'e
approximant\cite{buhler} (antiferromagnetic case) and the two-point
Pad\'e approximant\cite{baker} (ferromagnetic case). In \S IV we
present our conclusions. In appendix A we show that the trace of
powers of $S^z$ can be written as an expansion in the square of the
norm of the spin at each site. Finally, in appendix B we present the
high-temperature expansion, up to order $\beta^ 6$, of the Helmholtz
free energy of the spin-$s$ $XXZ$ model ($|\vec{s}|=1$), with a
single-ion anisotropy term and in the presence of an external magnetic
field.

\section{The high temperature behavior of the quantum 
spin-$s$ $XXZ$ chain}

We intend to study the thermodynamics of the Hamiltonian (\ref{hamt1})
for arbitrary spin, including its classical limit ($S \rightarrow
\infty$). In this limit, the thermodynamic functions diverge and to
render them finite we define a rescaled spin operator ${\bf s} \equiv
{\bf S}/\sqrt{S(S+1)}$. This rescaled spin operator has unitary norm
for all values of $S$. Rewriting the Hamiltonian \eqref{hamt1} in
terms of $s$ and redefining the parameters $J\equiv S(S+1)J'$,
$h\equiv \sqrt{S(S+1)}h'$ and $D\equiv S(S+1)D'$, we obtain

\begin{align}\label{hamiltonian}
{\bf H} = \sum_{i=1}^N J \left( {\bf s}_{i}, {\bf s}_{i+1}\right)_{\Delta} 
       - h s^{z}_{i} + D (s^{z}_{i})^2.    
\end{align}

\noindent This also describes the dynamics of the $XXZ$ model of a
spin of unitary norm and $(2S+1)$ $z$-components. We continue to use
the notation: $( {\bf s}_{i}, {\bf s}_{i+1})_{\Delta} \equiv s^{x}_{i}
s^{x}_{i+1} + s^{y}_{i} s^{y}_{i+1} + \Delta s^{z}_{i} s^{z}_{i+1}$.
Here, $s^x_i$, $s^y_i$ and $s^z_i$ stand for the spin-$s$ ($s=1/2, 1,
3/2, \cdots$) rescaled matrices in the {\it i}-th site of the chain;
$N$ is the number of sites in the periodic chain; and $\Delta$ is the
anisotropy constant in the $z$-direction.

In order to evaluate the Helmholtz free energy for arbitrary 
values of spin,  we first calculate 
its expansion for a set of spin values using 
the Hamiltonian (\ref{hamiltonian}).
We apply the method of Ref.\cite{chain_m} to calculate  those
$\beta$-expansions  for arbitrary values of  $J$, $\Delta$, $h$ and $D$.
We presented elsewhere\cite{jpcm03} the  Helmholtz free energy of 
Hamiltonian (\ref{hamt1})  for a number of spin values (semi-integer
and integer) up to $S=4$.  We use   the interpolation method
to extend our results for fixed values of $S$ to arbitrary spin value. 
For the sake of obtaining the $\beta$-expansion of
any spin-$S$ quantum chain up to order $\beta^6$, having the 
expansions presented in Ref. \cite{jpcm03} is not enough.

In order to derive the expansion of the Helmholtz free energy 
of the hamiltonian (\ref{hamt1})
up to order $\beta^n$ (for a survey of the method\cite{chain_m}, 
cf. section 2 of Ref. \cite{jpcm03}),
we have to calculate normalized
traces, at each site, of the products of $m_i$ 
 matrices  from the  set \{$S^x$,  $S^y$, $S^z$\}, 
 so that $m_i\le 2(k+1)$ and 
$k$ ranges over all possible powers of $\beta$, $k=1, 2, \cdots, n$.  
Due to the commutation relation 
$[S^x,S^y]= i S^z$ and to the fact that the products $S^xS^y$ and $S^yS^x$ can be
written in terms of $S^z$, $(S^z)^2$  and the identity matrix ${\bf 1}$, 
then the normalized trace at each site  is reduced to the calculation of traces of 
$(S^z)^l$, with $l \in N$.
 
In eq.\eqref{sn_nat} we show that the ${\rm tr}((S^z)^l)$, for even
values of $l$, is a polynomial of degree $(l+1)$ in $S$. Therefore the
polynomial in $S$ of highest degree at order $\beta^n$ is of order
$2(n+1)$. In order to carry out the interpolation of the expansion up
to order $\beta^6$, we calculated the $\beta$-expansion of the
Helmholtz free energy from the spin $S= 1/2$ up to $S=7$. Later, by
inspection of the expansion in $S$ and in $\beta$, we verified that
the expansion in $S$ can be rewritten in terms of $S(S+1)$, that is,
in terms of the eigenvalue of a constant of motion (the square of the
norm of the spin-${\bf S}$ at each site). In appendix \ref{appendixA}
we show, for arbitrary value of spin, that $tr[(S_i^z)^{2l}]$, $l= 1,
2 \cdots $, is a polynomial of degree $(l-1)$ in the parameter
$S(S+1)$.

The thermodynamic properties of the quantum 
spin-$S$ $XXZ$ chain, in
the high temperature region, can be obtained 
from its Helmholtz free energy. We present 
in appendix B the coefficients
of the Helmholtz free energy ${\mathcal W}_s(\beta)$ derived 
from the Hamiltonian (\ref{hamiltonian}),
up to order $\beta^6$, for any value of 
$S (S = 1/2, 1, 3/2, \cdots)$. In particular, 
the result of appendix B is also valid at the isotropic point of the
Hamiltonian \eqref{hamiltonian} with $D=0$ and $h=0$, when
the model has rotational symmetry.  The $\beta$-expansion of the
Helmholtz free energy of Hamiltonian (\ref{hamt1}) is easily
obtained from result (\ref{free_e}) (see appendix \ref{B}).
Fukushima {\it et al.}\cite{fukushima} obtained the
$\beta$-expansion of the specific heat and magnetic
 susceptibility  per site  of  the latter model for
arbitrary spin at the isotropic point and in the absence of
an external magnetic field. Our results
agree with theirs,  up to order $\beta^6$.

The range of validity in $\beta$ of our high-temperature
expansions obviously depends on the particular values given to the parameters
of \eqref{hamiltonian}. The 
thermodynamic  properties of the chain depend on the sign of 
the product $J\Delta$. We let  $J>0$ and let $\Delta$ 
refer to either the ferromagnetic   $(\Delta<0)$ or 
the antiferromagnetic $(\Delta>0)$ phases.

\subsection{Comparison with  known classical limits ($S\rightarrow\infty$)}

An interesting check  to our expansion \eqref{free_e} is to
 verify if it recovers known
classical limits ($S\rightarrow\infty$ or $y\rightarrow 0$,
where $y= \frac{1}{S(S+1)}$) of the $XXZ$ chain.
The  classical limit of the $XXZ$ chain of the Hamiltonian 
(\ref{hamiltonian})  with $h=0$ and $D=0$ has
been solved by Joyce\cite{joyce_prl}; this model (including 
the anisotropic constant $\Delta$ in the 
Hamiltonian \eqref{hamiltonian}) is exactly solvable 
and its Helmholtz free energy is given
by the radial component of the spheroidal wave function

\begin{align}\label{f_class}
{\mathcal W}_{\infty} ( \mu; \beta)=-\frac{1}{\beta}\ln \Big({R}_{0,0}(-i \beta
J/\sinh\mu,\cosh\mu)\Big) ,
\end{align}

\noindent where $\mu=\tanh^{-1}(\frac{1}{\Delta})$ and
$R_{0,0}$ is the first radial spheroidal  wave
function\cite{abramowitz}.  For $|\Delta|=1$,  eq. \eqref{f_class}
becomes the well-known result obtained by Fisher\cite{fisher},
whose explicit expression is 
${\mathcal W}_{\infty} ( \pm\frac{ \pi}{2}, \beta)=
-\frac{1}{\beta}\ln[\sinh(\beta J)/(\beta J)]$.

The classical limit of the $\beta$-expansion \eqref{free_e} is
obtained by letting $y=0$.  In order to have a shorter $\beta$-expansion 
of the Helmholtz  free energy at $y=0$, we take  $D=0$, and obtain

\begin{eqnarray}\label{w_clss}
{\mathcal W}_{\infty}(\Delta, \tilde{h}, 0; J\beta)&=&
\left( - {\displaystyle \frac {1}{6}} 
\,\tilde{h}^{2} 
- {\displaystyle \frac {1}{9}} 
- {\displaystyle \frac {1}{18}} \,\Delta ^{2}\right)\, (J\beta) 
+ {\displaystyle \frac {1}{9}} \,\Delta \,\tilde{h}^{2}\, (J\beta) ^{2}  
+ \left({\displaystyle \frac {1}{180}} \,\tilde{h}^{4} 
- {\displaystyle \frac {1}{1350}}  
- {\displaystyle \frac {7}{135}} \,\Delta ^{2}\,\tilde{h}^{2} \right. \nonumber\\
& &\left. - {\displaystyle \frac {7}{2700}} \,\Delta^{4} 
 + {\displaystyle \frac {2}{135}} \,\tilde{h}^{2} 
 + {\displaystyle \frac {2}{225}} \,\Delta ^{2}\right)\, (J\beta) ^{3}   
+ \left( - {\displaystyle \frac {2}{135}} \,\Delta \,\tilde{h}^{2} 
- {\displaystyle \frac {4}{225}} \,\Delta  
+ {\displaystyle \frac {46}{2025}} \,\Delta ^{3}\right)\,\tilde{h}^{2} (J\beta) ^{4} \nonumber\\
& & + \left({\displaystyle \frac {424}{42525}} \,\Delta ^{2}\,\tilde{h}^{2}  
 + {\displaystyle \frac {179}{9450}} \,\Delta ^{2}\,\tilde{h}^{4} 
 - {\displaystyle \frac {359}{42525}} \,\Delta ^{4}\,\tilde{h}^{2} 
 - {\displaystyle \frac {2}{2835}} \,\tilde{h}^{2} 
 - {\displaystyle \frac {1}{525}} \,\tilde{h}^{4} 
 - {\displaystyle \frac {107}{2679075}} \,\Delta ^{6} \right. \nonumber\\
& &  \left. + {\displaystyle \frac {212}{893025}} \,\Delta ^{4}
 - {\displaystyle \frac {632}{893025}} \,\Delta ^{2} 
 + {\displaystyle \frac {422}{2679075}}  
 - {\displaystyle \frac {1}{2835}} \,\tilde{h}^{6}\right) (J\beta) ^{5}\mbox{} 
 + \left({\displaystyle \frac {16}{2835}} \,\Delta \,\tilde{h}^{2}  
 - {\displaystyle \frac {248}{14175}} \,\Delta ^{3}\,\tilde{h}^{2} \right. \nonumber\\
& &  \left. + {\displaystyle \frac {2566}{893025}} \,\Delta ^{5}
 + {\displaystyle \frac {1264}{893025}} \,\Delta  
 - {\displaystyle \frac {836}{178605}} \,\Delta ^{3}   
+ {\displaystyle \frac {1}{525}} \,\Delta \,\tilde{h}^{4}\right)\tilde{h}^{2} (J\beta) ^{6}
+\mathcal{O}((J\beta)^7),
\end{eqnarray}

\noindent where $\tilde{h} \equiv h/J$.

From (\ref{w_clss}) we see that in the absence of 
the  external magnetic field ($h=0$) the  Helmholtz free energy  is an even 
 function of  the anisotropic parameter $\Delta$. Under this condition,
 the classical limit of the 
$XXZ$  model does not  distinguish between
the  ferromagnetic and the antiferromagnetic phases in the region of high 
temperatures.

\begin{figure}[h!t]
\begin{center}
\includegraphics[width=9cm,height=14cm,angle=-90]{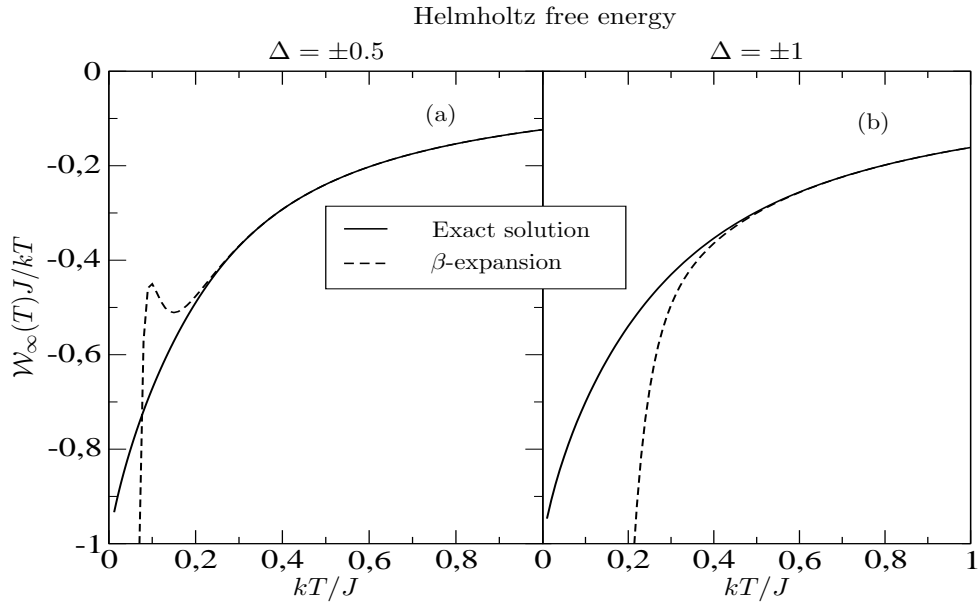}
\caption[fig_1]{The Helmholtz free 
energy ${\mathcal W}_{\infty}$ (for $S\rightarrow \infty$)  per
unit of $kT/J$ as a function of $kT/J$ for the classical $XXZ$ model
in the absence of the single-ion anisotropy term ($D=0$) and an
external magnetic field ($h=0$). In ($a$) we plot the anisotropic
cases with $\Delta=\pm 0.5$ and in 
($b$) the isotropic cases
($\Delta=\pm 1$). The solid line 
refers to the numerical solution of
(\ref{f_class}) and the dashed line 
to the expansion (\ref{w_clss}).}
\label{fig_1}
\end{center}
\end{figure}

Fig. (\ref{fig_1}$a$) shows a comparison of our $\beta$-expansion of
the Helmholtz free energy to the numerical solution of
eq.\eqref{f_class}, for $h=0$, $D=0$ and $\Delta= \pm 0.5$. They are
in good agreement in the high temperature region; the percental error
between them at $kT/J =0.2$ is less than 2.8\%.
Fig. (\ref{fig_1}$b$) performs the same comparison, but at the
isotropic point ($\Delta= \pm 1$). We also have good
agreement, but for higher temperatures;  percental error
is about $3\%$ for $kT/J=0.4$.

In conclusion, the classical limit of  the expansion \eqref{free_e} 
(for $y=0$)  coincides with Joyce's solution \eqref{f_class}  of the
Helmholtz free energy of the Hamiltonian (\ref{hamiltonian}) in a 
larger interval of temperature 
than that which we call ``high temperature region''.

\subsection{The thermodynamics of the  spin-$s$ $XXZ$ model}

The greater the (finite) spin, the greater the number of degrees of
freedom to be handled, thus the more involved it gets to compute
thermodynamic properties of the quantum spin-${\bf S}$ model. In
contrast, for infinite values of spin, the calculation of those
properties for the classical version of the model is simpler, since we
can treat the spin as a vector that rotates continuously. It is
interesting to know for which values of $S$ the quantum spin model can
be well approximated by its classical version, at least in the high
temperature region. 
  
In Ref. \cite{jpcm03} we compared various thermodynamic functions of
the chain model at the Heisenberg point ($\Delta = \pm 1$), in the
absence of the single-ion anisotropy term and external magnetic
fields. We verified, in the high temperature region, that
thermodynamic functions like the magnetic susceptibility per site have
a classical behavior, for $S \geq 2$. Our intention in this section is
to extend the results of Ref. \cite{jpcm03} in order to include in
this comparison (quantum $\times$ classical behavior) of spin models
with anisotropies in the $z$-direction ($\Delta \not= \pm 1$ and $
D\not= 0$) and in the presence of an external magnetic field. This
analysis can be realized from the expansion (\ref{free_e}) of
${\mathcal W}_s (\beta)$. The referred thermodynamic functions are all
derived from the Hamiltonian (\ref{hamiltonian}).

Let $C_s(\beta)$ and $C_S(\beta)$ be the specific heat per site for
the spin-$s$ model (with unitary norm) and for the spin-$S$
chain (with norm equal to $\sqrt{S(S+1)}$), respectively. The specific
heat per site is calculated from a derivative of the Helmholtz free
energy ($C (\beta) = - \beta^2 \frac{\partial ^2 (\beta {\mathcal
W}(\beta))}{\partial \beta^2}$). From the results (\ref{b2}) we
obtain,
 
 \begin{subequations}\label{5}
 
 \begin{eqnarray}
 C_s (J, \Delta, h, D; \beta) & = & C_S \left(J, \Delta, \sqrt{S(S+1)}  h, D; 
\frac{\beta}{S(S+1)}\right)
                                 \label{5a} \\
 %
 %
 & = & C_S \left( \frac{J}{S(S+1)}, \Delta, \frac{h}{\sqrt{S(S+1)}}, \frac{D}{S(S+1)}; 
\beta\right).
                            \label{5b}
 \end{eqnarray}
 
 \end{subequations}
 
\noindent These results show that $C_S$ is a homogeneous function of
zero degree for all temperatures. The $\beta$- expansion of the
specific heat derived from eq. (\ref{free_e}) satisfies eqs.(\ref{5})
identically.

From the high temperature expansion of $C_s $ we verify that
 $C_s =   - (\frac{D^2}{15 S(S+1)} - \frac{h^2}{3} - \frac{2}{9} -
 \frac{4 D^2}{45} - \frac{\Delta ^2}{9} ) \beta^2  + {\cal O} (\beta^3)$.
 In the high temperature region, the  $XXZ$ model presents
  a tail of the Schottky peak\cite{schottky}
 ($C_{Sch} \propto \beta^2$), for all values of $S$.

\begin{figure}[h!t]
\begin{center}

\includegraphics[width=10cm,height=15cm,angle=-90]{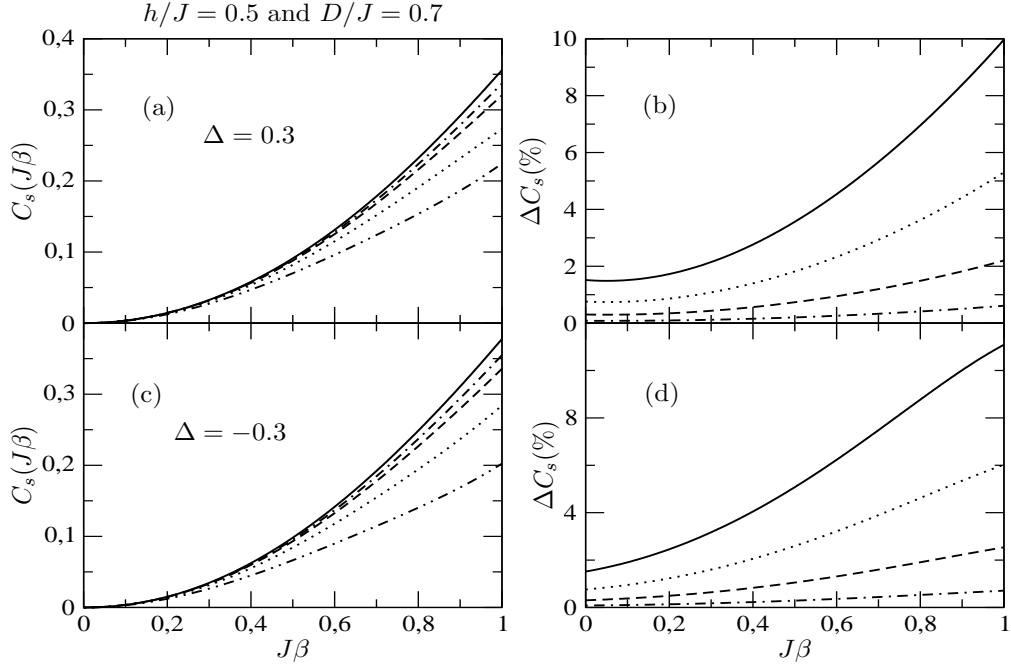}

\caption[fig_2]{Figs. $(a)$ and $(c)$ show the specific heat
$C_s(J\beta)$; $(b)$ and $(d)$ show the relative percental errors 
between $C_s(\beta)$ and $C_{\infty}(\beta)$ (the classical limit) in 
$(a)$ and $(c)$, respectively. 
For all plots we have $h/J= 0.5$ and  $D/J= 0.7$. 
In $(a)$ we let $\Delta= 0.3$ and in $(c)$ we have $\Delta = - 0.3$. 
In $(a)$ and $(c)$, line styles correspond to different values of spin, as follows: 
$S=1/2$ (dashed and double-dotted lines), $S=1$ (dotted line),
$S=2$ (dashed line), $S=3$ (dashed and dotted lines) and $S\rightarrow\infty$ 
(solid line). In $(b)$ and $(d)$ we have $S=2$ (solid line),
$S=3$ (dotted line), $S=5$ (dashed line) and $ S=10$ (dashed and dotted line).}
\label{fig_2}
\end{center}
\end{figure}

Fig. \ref{fig_2} shows the specific heat per site as a function of $J\beta$
for distinct values of $S$, including the classical limit ($S\rightarrow \infty$)
of the Hamiltonian (\ref{hamiltonian})  and the 
relative percental error of this 
function for various values of $S$ in relation to the classical specific 
heat per site. We take $h/J= 0.5$ and $D/J = 0.7$. 
Fig. (\ref{fig_2}a) pictures the antiferromagnetic case ($\Delta= 0.3$)
and (\ref{fig_2}c) the ferromagnetic case ($\Delta=-0.3$).
Figs. (\ref{fig_2}b)  and (\ref{fig_2}d) show the
  relative percental error, i.e., 
$\left|\frac{C_{\infty}- C_s}{C_{\infty}}\right|\times 100\%$,
of Figs. (\ref{fig_2}a)  and (\ref{fig_2}c), respectively, where
$C_{\infty}$ is the classical limit of the specific heat per site.  
In both cases,  even the $S=2$ model 
 behaves classically (within an error 
smaller than $2\%$) up to $J\beta = 0.3$. Within this range of 
error,  we can also neglect the
quantum nature of the spin model with $S\geq 5$  
up to $J\beta\sim 1$.

Let $\chi_s (\beta)$ and $\chi_S (\beta)$ be the magnetic
susceptibilities per site of the spin-$s$ model (with unitary norm) and 
of the spin-$S$ model (with norm equal to $\sqrt{S(S+1)}$), respectively.
The relation between the magnetic susceptibilty per site and 
the Helmholtz free  energy is 
$\chi (\beta) = - \frac{\partial^2{\mathcal W}(\beta)}{\partial h^2}$.
From the results (\ref{b2}) we obtain the relation between $\chi_s$ and
$\chi_S$, 

\begin{subequations} \label{6}

\begin{eqnarray}
\chi_s (J, \Delta, h, D; \beta) & = & 
     \chi_S \left(J, \Delta, \sqrt{S(S+1)}  h, D; \frac{\beta}{S(S+1)}\right)
                                 \label{6a} \\
 %
 %
 & = &  \frac{1}{S(S+1)} 
 \chi_S \left( \frac{J}{S(S+1)}, \Delta, \frac{h}{\sqrt{S(S+1)}}, 
 \frac{D}{S(S+1)}; \beta\right).
                            \label{6b}
\end{eqnarray}

\end{subequations}

\noindent Equating the r.h.s. of eqs. (\ref{6a}) and (\ref{6b})
we obtain that $\chi_S$ is a homogenous function of degree 1 
for all values of temperature.
Its $\beta$-expansion satisfies this property.

\begin{figure}[h!t]
\begin{center}
\includegraphics[width=10cm,height=15cm,angle=-90]{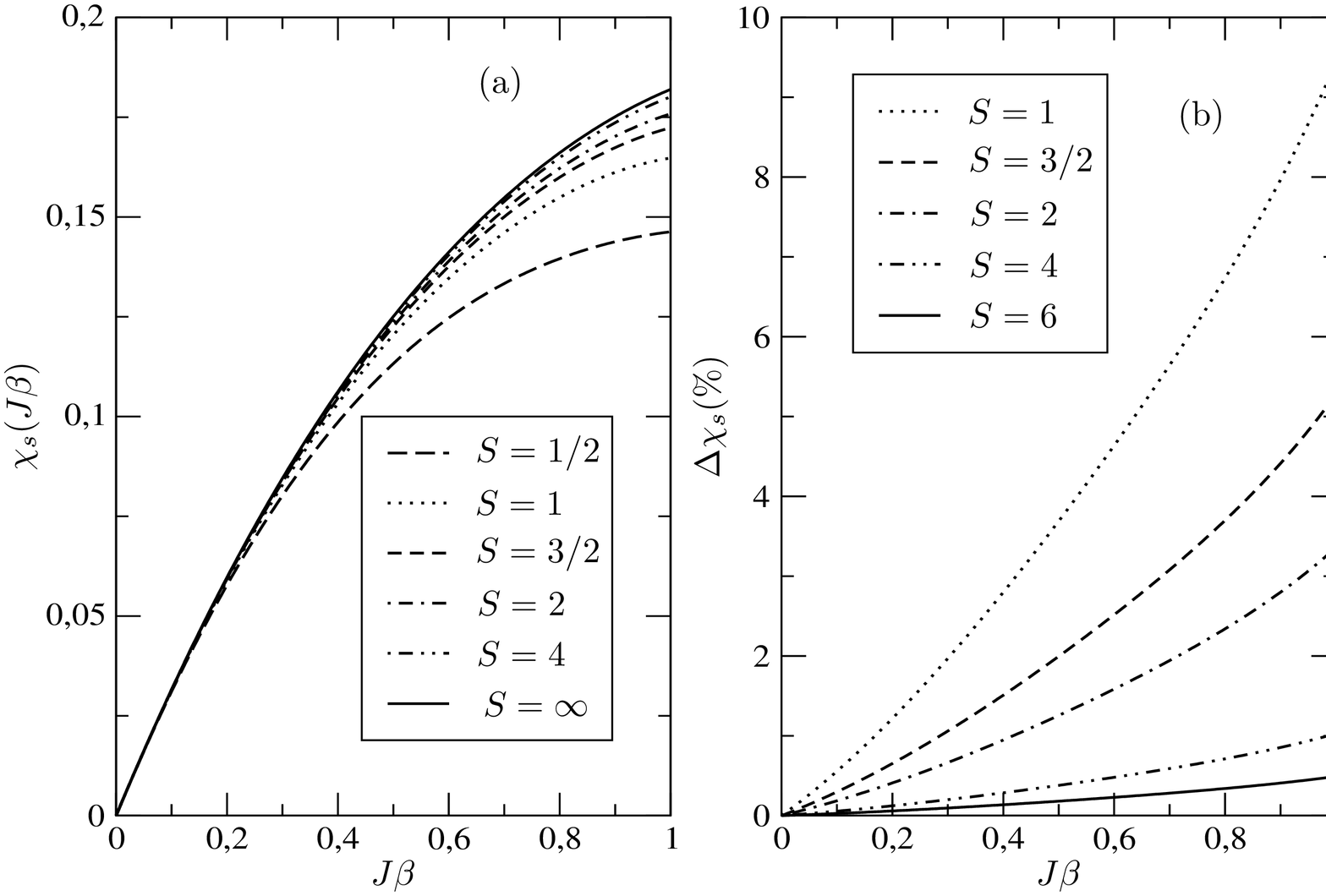}

\caption[fig_3]{$(a)$ The magnetic susceptibility
per site $\chi_s(J\beta)$ obtained from expansion (\ref{free_e})
 for various values of spin, including the classical limit (solid line). 
 We take $\Delta=1$,  $h/J= 0.3$ and $D/J= -0.5$.  
 $(b)$ Relative percental error between $\chi_s$ and $\chi_{\infty}$,
  for several values of $S$.}
\label{fig_3}
\end{center}
\end{figure}

Fig. \ref{fig_3} shows the magnetic susceptibility 
versus $J\beta$ for various
finite values of $S$ and the classical chain model 
($S\rightarrow \infty$).
Fig. (\ref{fig_3}b) shows the relative percentage error of the magnetic 
susceptibilty with respect to the classical curve  
for $S=1, 3/2, 4$ and $6$. In Fig. \ref{fig_3} we have
$\Delta=1$, $h/J= 0.3$ and $D/J = 0.5$. 
From Fig. (\ref{fig_3}b) 
we verify that the magnetic susceptibility of the
spin-3/2 can be approximated by the classical result up to
$J\beta \sim 0.5$ with an error smaller than $2\%$. The classical magnetic
susceptibility curve is a good approximation of the spin-2 model up
to $J\beta \sim 0.7$.

\begin{figure}[h!t]
\begin{center}

\includegraphics[width=10cm,height=15cm,angle=-90]{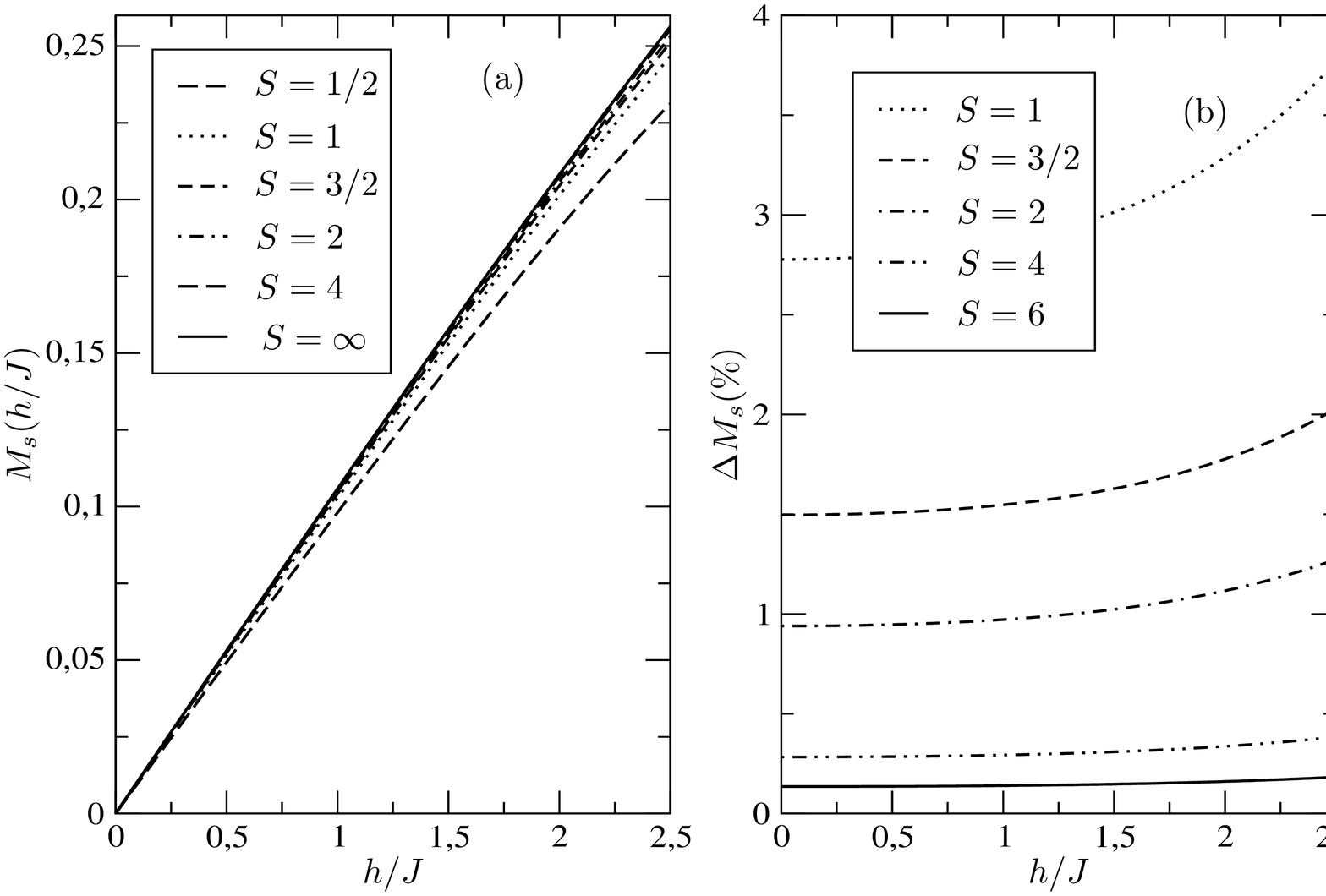}
\caption[fig_4]{$(a)$ Magnetization $M_s(h/J)$ 
at $J\beta= 0.4$,  calculated  from  the
$\beta$-expansion of ${\mathcal W}_s (\beta)$,
 for various values of spin, including the classical model  ($S\rightarrow \infty$). We
take $\Delta=1$  and $D/J= -0.5$.  $(b)$ Relative
percental error of $M_s$, for several values of $S$, with respect to 
$M_{\infty}$.}
\label{fig_4}
\end{center}
\end{figure}

Fig. \ref{fig_4} shows the magnetization of the $XXZ$  model
of the spin with unitary norm versus 
$h/J$ ($M(\beta) = - \frac{{\mathcal W}(\beta)}{\partial h}$). We call
$M_s (\beta)$  the  magnetization derived from Hamiltonian (\ref{hamiltonian})
and $M_S (\beta)$ the magnetization derived from (\ref{hamt1}).  From 
eqs. (\ref{b2}), we also obtain 

\begin{subequations}  \label{7}

\begin{eqnarray}
M_s (J, \Delta, h, D; \beta) & = & \sqrt{S(S+1)} \;
M_S \left(J, \Delta, \sqrt{S(S+1)}  h, D; \frac{\beta}{S(S+1)}\right)
                                 \label{7a} \\
 %
 %
 & = & \frac{1}{\sqrt{S(S+1)}}   \;
 M_S \left( \frac{J}{S(S+1)}, \Delta, \frac{h}{\sqrt{S(S+1)}}, 
 \frac{D}{S(S+1)}; \beta\right).
                            \label{7b}
\end{eqnarray}

\end{subequations}

\noindent From eqs. (\ref{7}) we obtain that $M_S $ is 
a homogeneous function 
of degree 1 at all temperatures. This condition is satisfied by the 
$\beta$-expansion of  the magnetization, derived  from eq. (\ref{free_e}).

Fig. (\ref{fig_4}a) shows the magnetization $M_s$ as a function
of $h/J$ at $J\beta= 0.4$. We choose the same values of constants
in the Hamiltonian (\ref{hamiltonian}) as in Fig. (\ref{fig_3}),  
$\Delta=1$ and $D/J = -0.5$.  Comparison of 
Figs. (\ref{fig_3}b) and (\ref{fig_4}b) shows a closer
similarity of quantum and classical magnetization curves
(even for low values of spin such as $S=3/2$)
than that of magnetic susceptibility
curves, as far as high temperatures are concerned.

\begin{figure}[h!t]
\begin{center}
\includegraphics[width=10cm,height=15cm,angle=-90]{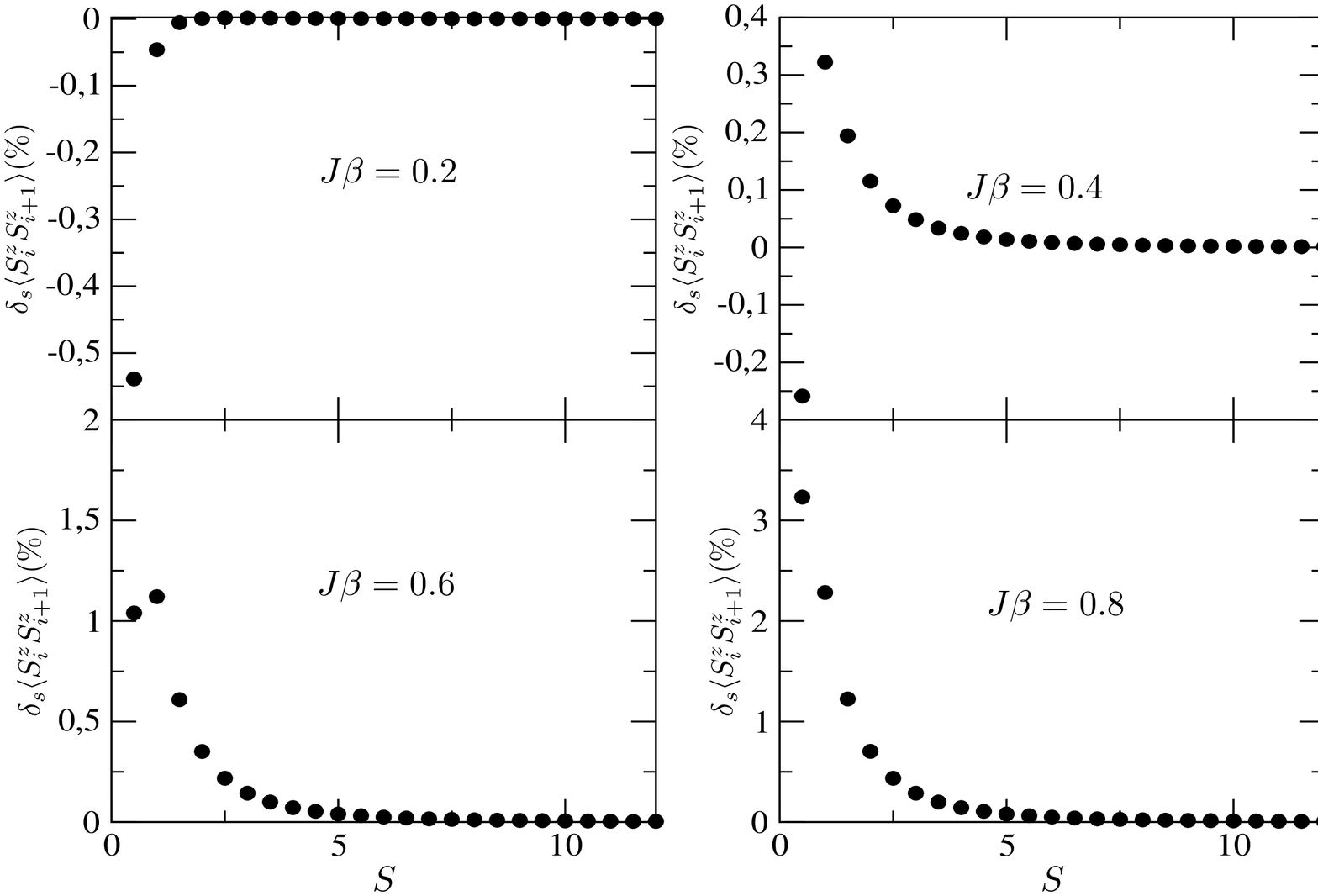}
\caption[fig_5]{
Percental variation of the correlation function
$\delta_s \langle S_i^z S_{i+1}^z\rangle$ as a function of $S$,  for
several values of $J\beta$ in the high temperature region.
In all plots we take $\Delta =1$, $h/J = 0.3$ and
$D =  -0.5$. }
\label{fig_5}
\end{center}
\end{figure}

The correlation function of spin $z$-components between first
neighbors is written as ${\langle S_i^z S_{i+1}^z\rangle_s} =
\frac{\partial {\mathcal W}_s (\beta)}{ \partial \Delta}$, for a given
spin $s$. Its percental variation (when $s$ varies by half-integer
steps), shown in Fig. (\ref{fig_5}), is defined as

\begin{equation}
\delta_s \langle S_i^z S_{i+1}^z\rangle\equiv 
\frac{\langle S_i^z S_{i+1}^z\rangle_{(s+1/2)}
 - \langle S_i^z S_{i+1}^z\rangle_s}
{\langle S_i^z S_{i+1}^z\rangle_s}.
\end{equation}

\noindent For $S\gg 1$, the function $\delta_s \langle S_i^z
S_{i+1}^z\rangle$ has an expansion in $S^{-1}$ whose leading term is
$S^{-3}$, and its coefficient depends on the temperature. In Fig.
(\ref{fig_5}) we take $\Delta=1$, $h/J= 0.3$ and $D/J= -0.5$. In the
very high temperature of $J\beta =0.2$, only the correlation functions
for $S=1/2$ and $S=1$ differ from the classical curve by a difference
smaller than $0.6\%$. As we lower the temperature, spins of higher
values have correlation functions between first neighbors that differ
slightly from the classical ones, but even at $J\beta\sim 0.8$, the
correlation function for $S=2$ approximates the classical result with
an error smaller than $1\%$.

\section{Extension  of the Thermodynamics of the quantum spin-$s$  $XXZ$ chain 
to Lower Temperatures}

Although there are only seven terms in the high-temperature expansion
of the Helmholtz free energy of the spin-$s$ $XXZ$ model (see
appendix (\ref{B})), different Pad\'e approximants permit us to
enhance our analytic results and extend them to lower temperatures. In
this section, we shall assume a quantum spin chain with unitary norm.

Bernu and Misguich \cite{bernu} presented an approach to
``interpolate'' the high- and low-temperature behaviors of the
specific heat per site, provided that the ground state energy per site
of the chain is known. We refer the reader to their work for details
on the method.

The  $XXZ$ models (with $S \not= 1/2$), at the isotropic 
points ($\Delta \pm 1$),  are among the most 
studied ones. For this reason, in this section
we calculate the Pad\'e approximants at those points, for the sake
of comparison of our results to the literature.

At $T\sim 0$ the behavior of the specific heat per site is distinct  for 
gap and gapless models. Ferromagnetic and  half-odd integer spin
 chains do not have excitation  energy gaps and the specific 
 heat per site around $T=0$ is $C_s \sim T^{p/q}$, where
 $p$ and $q$ are integers\cite{bernu}.   By contrast, the integer 
 antiferromagnetic spin chains  have an energy gap between 
 the ground state energy and the lowest exited
state  and this thermodynamic  function,  in the region of  $T\sim 0$,
 is $C_s \sim  e^{-\frac{E_{gap}}{kT}} T^{\alpha}$\cite{haldane}. 
As the spin increases, this  energy gap decreases rapidly;  
it only vanishes at the classical limit ($S\rightarrow \infty$), though.

In Ref. \cite{yamamoto96}, Yamamoto studied
numerically   the thermodynamics  of the $S=2$  $XXZ$ model  
with $\Delta=1$, $D=0$  and  $h=0$.
  The ground state energy per site of this model for
 the non-unitary spin chain was obtained in 
 Ref.  \cite{todo01}; 
 the  ground state energy per site  of the unitary 
 spin chain can then be easily obtained, being equal to $e_0/J= -0.79354$.  
 This  model exhibits 
 an  energy gap $E_{gap}/J\approx  0.0593$ \cite{todo01}. 
 At very low  temperature it  is expected that the
 specific heat vanish  exponentially  ($C_s\sim \exp(-E_{gap}/kT)$) 
  as $T\rightarrow 0$. On the other hand, the spin wave  theory 
gives a  linear behavior for the specific heat 
over a finite interval of intermediate low temperatures \cite{kubo}.
In order to enhance our $\beta$-expansion of $C_2 (T)$ to 
lower temperatures,  we simply use  $C_2\sim T$.
The entropy at low temperatures behaves as 
${\mathcal S}(e)=(e-e_0)^{1/2}$, where $e$ is the ground state energy per site, and it 
 has an essential singularity when $e\rightarrow e_0$. 
 Following Ref. \cite{bernu}, we construct an auxiliary function 
${\mathcal G}(e) \equiv ({\mathcal S}(e))^2$,  which is analytic in
the interval $[e_0,0]$, and use the Pad\'e approximant to
 fit ${\mathcal G}(e)$, so the entropy becomes ${\mathcal
S}(e)=\tilde{\mathcal G}(e)^{1/2}$. By $\tilde{\mathcal G}(e)$ 
we denote the Pad\'e approximant  realized over ${\mathcal G}(e)$.
The specific heat per site of the model is obtained using the relation  
$C_2(e)= -{\mathcal S}'(e)^2/{\mathcal S}''(e)$,
which can be plotted in 
 parametric form $\{T(e),C_2(e)\}$  (for details see Ref. \cite{bernu}). 

\begin{figure}[h!t]
\begin{center}
\includegraphics[width=6cm,height=8.5cm,angle=-90]{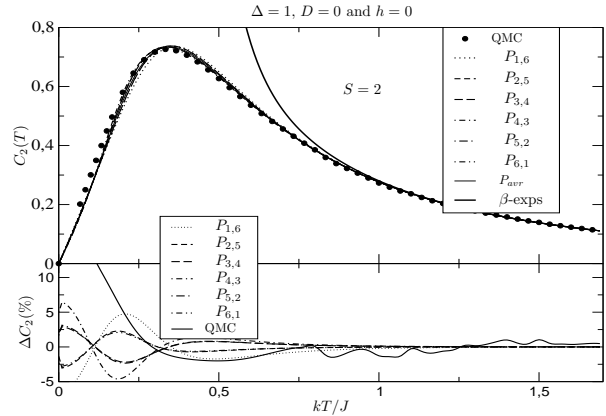}
\caption[fig_6]{(Top) The Pad\'e approximant of the specific heat as a
function of $kT/J$, for $S=2$ with $\Delta=1$, $D=0$ and $h=0$.
The thicker continuous line stands for the $\beta$-expansion of
$C_2 (T)$.    (Bottom) The $\Delta C_2 (T)100\%$
 represents the percental deviation of
all acceptable (non-singular) Pad\'e approximants and the QMC
results\cite{yamamoto96}
with respect to the average of Pad\'e  approximants.}
\label{fig_6}
\end{center}
\end{figure}

In Fig. (\ref{fig_6}) (top) we compare all ``acceptable'' Pad\'e
approximants
$P_{n,m}$ (i.e., all approximants so that $n+m=7$ and that do not possess
spurious
singularities) and their average $P_{avr}$, to the numerical results
obtained by
the quantum Monte Carlo method (QMC)\cite{yamamoto96}
and the $\beta$-expansion of $C_2(T)$.
Fig. (\ref{fig_6}) (bottom) displays the percental deviation
of each Pad\'e approximant  $P_{m,n}$ with respect to
$P_{avr}$. We define
$\Delta C_2 (\%) = (P_{m,n}-P_{avr})/P_{avr} \times 100\%$.
The  high temperature expansion satisfactorily describes  $C_2 (T)$ up
to $kT/J \sim 0.9$.  The solid line corresponds to
the percental deviation of $P_{avr}$ with respect to the QMC prediction.
Although we have only seven terms in expansion (\ref{free_e}) of the
Helmholtz free energy, the percental relative difference
of the Pad\'e approximants is smaller than
 $2,5\%$ for $kT/J \thickapprox  0.25$, allowing good precision
  in calculating the temperature where $C_2(T)$ is maximum.

\begin{figure}[h!t]
\begin{center}
\includegraphics[width=6cm,height=8.5cm,angle=-90]{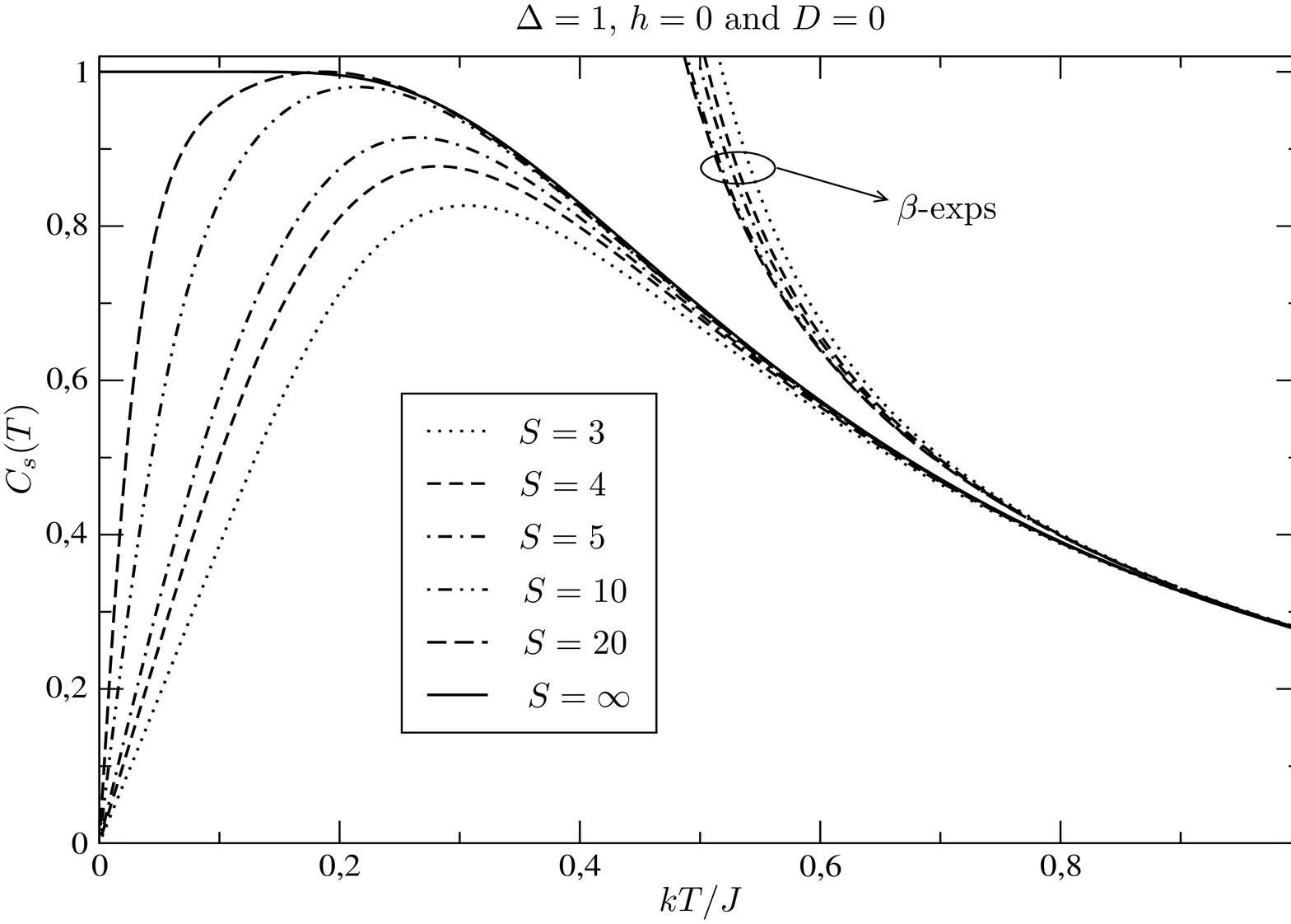}
\caption[fig_7]{ The specific heat versus temperature for several values
 of spin at the isotropic  point $\Delta= 1$   with $h=0$ and $D=0$.
 For each value of spin,   the curve is obtained by   taking the 
 average of all ``acceptable'' Pad\'e approximants (i.e., with no spurious singularities), 
 whereas in the classical limit corresponds to Fisher's solution.}
\label{fig_7}
\end{center}
\end{figure}

\begin{figure}[h!t]
\begin{center}
\includegraphics[width=6cm,height=8.5cm,angle=-90]{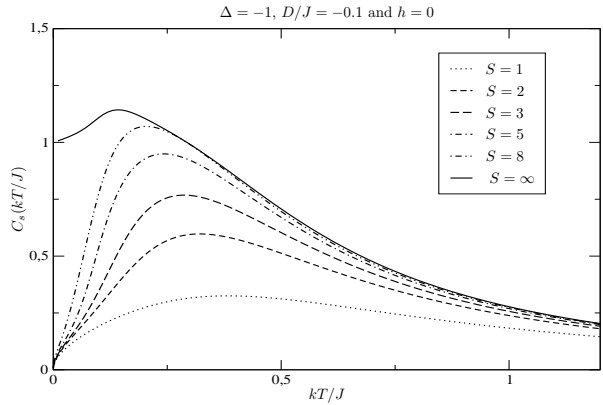}
\caption[fig_9]{The specific heat per site versus temperature 
($kT/J$) for several values of spin in the  ferromagnetic
case. We take:  $\Delta=-1$ with $h=0$ and $D=-0.1$.
Each  curve is obtained by plotting its  $P_{avr}$, while the
classical limit is obtained numerically.}
\label{fig_9}
\end{center}
\end{figure}

Figs. (\ref{fig_7}) and (\ref{fig_9}) show the specific heat for
several values of spin for the antiferromagnetic ($\Delta=1$) and
ferromagnetic ($\Delta= - 1$) cases, respectively, allowing us to
observe how the specific heat per site approaches its classical limit
as $S$ varies.  In order to show how the Pad\'e approximants enhance the
high temperature solution of the specific heat per site, Fig.
(\ref{fig_7}) also displays the $\beta$-expansion of each $C_s(T)$.
Following the spin wave theory, we assume that the specific heat
behaves as $C_s\sim T$ for the antiferromagnetic\cite{kubo} case, and
as $C_s\sim T^{1/2}$ in the ferromagnetic\cite{takahashi} case. 
We fit the numerical values of the ground state energy per site
calculated in Ref. \cite{todo01}, for the antiferromagnetic case with
$S= 1, 2$ and $3$, and extrapolate them to the integer
antiferromagnetic spin $S$ chain. Our approximate result for this
ground state energy per site is $e_0/J \thickapprox - (1
+0.3641/S+0.029/S^2+0.0086/S^3)S/(S+1)$. For the ferromagnetic case at
the Heisenberg point ($\Delta \pm 1$) with $h=0$ and $D=0$, the ground
state energy simply becomes $e_0/J = -S/(S+1)$.

Fig. (\ref{fig_7}) shows the specific heat as a function of
temperature for $\Delta= 1$, with $D=0$ and $h=0$. The curves
correspond to the $P_{avr}$ of this thermodynamic function for the
following set of spin values, as well as their respective
$\beta$-expansions: $S=\{3,4,5,10,20\}$.
 The exact classical curve
corresponds to Fisher's solution\cite{fisher}. From our numerical
results, we verify that even the Pad\'e approximants for $S=3$
deviates less than $2\%$ from the classical result ($S \rightarrow
\infty$) up to $kT/J \sim 0.64$, while for $S=5$ this deviation is
about $2\%$ only at $kT/J \sim 0.42$.

\begin{figure}[h!t]
\begin{center}
\includegraphics[width=6cm,height=8.5cm,angle=-90]{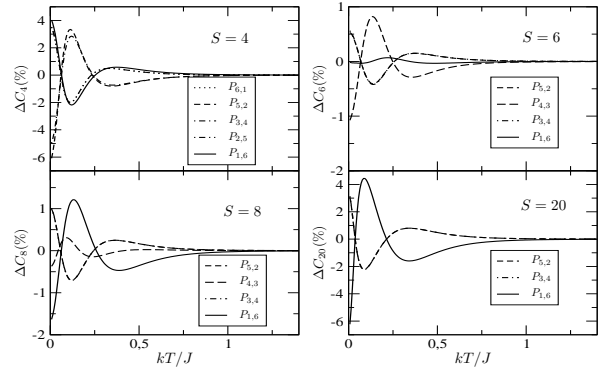}
\caption[fig_8]{The relative percental difference of all ``acceptable'' Pad\'e's
approximants to the average  
Pad\'e $P_{avr}$, for the specific heat per site for the antiferromagnetic
case ($\Delta = 1$ with $h=0$ and $D=0$), with $S=4, 6, 8$ and $20$.    
Approximants with spurious singularities have been excluded.}
\label{fig_8}
\end{center}
\end{figure}

Except for the classical limit ($S \rightarrow \infty$), to the best
of our knowledge there are no known results of the specific heat per
site for $S \geq 3$ for the whole range of temperature. In order to
assure the validity of our results for the specific heat at lower
temperatures, for distinct $S$ as presented in Fig. (\ref{fig_7}), we
plotted in Fig. \ref{fig_8} the relative percental difference between
the Pad\'e approximants without spurious poles and the average Pad\'e
for several values of spin ($S=4, 6, 8$ and $20$). 
For all those values of spin, it can be seen
that  deviations get larger than $2\%$
for $kT/J \lesssim 0.25$, whereas their respective
$\beta$-expansion are good approximations only
up to $kT/J \sim 0.8$.

In Fig. (\ref{fig_7}) we verify that the curves of the specific heat
for the antiferromagnetic case are close to the classical curve up to
low temperatures, even for $S=3$. Comparing its behavior to that of
the ferromagnetic case, we plot in Fig. (\ref{fig_9}) the specific
heat for several values of spin, taking $\Delta=-1$, $h=0$ and
$D=-0.1$. We assume that the specific heat for the ferromagnetic case
and small values of $D$ still behaves as $C_L\sim T^{1/2}$ and the
ground state energy per site is $e_0/J = - S/(S+1)-DS/(S+1)-D/3$. The
first term in the ground state energy corresponds to the case $D=0$;
the second one is the contribution of the single-ion anisotropy to the
ground state energy; and the last term corresponds to its global
shift. In order to simplify our calculations, we impose that the
average energy vanishes at $\beta=0$ (otherwise it would be equal to
$D/3$, that is, the value obtained from our high temperature
expansion).

\begin{figure}[h!t]
\begin{center}
\includegraphics[width=6cm,height=8.5cm,angle=-90]{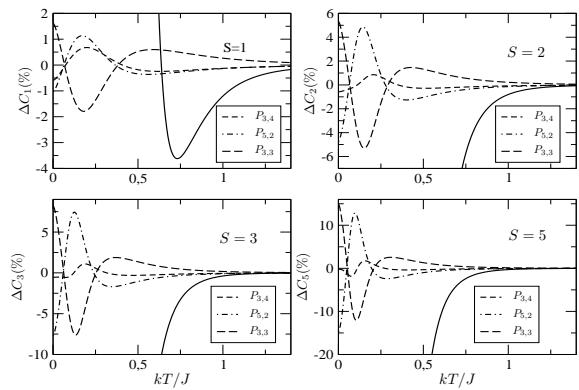}
\caption[fig_10]{The relative percental difference  of the Pad\'e's
approximants to the average  Pad\'e $P_{avr}$
of the specific heat per site for the antiferromagnetic
case ($\Delta = 1$ with $h=0$ and $D=0$),
for $S=4, 6, 8$ and $20$. Approximants with spurious singularities
have been excluded. The continuous line in each picture
represents the percental difference  between the $\beta$-expansion of
$C_s (T)$ and $P_{avr}$.}
\label{fig_10}
\end{center}
\end{figure}

As in the antiferromagnetic case, we do not know of results that would
allow comparison regarding the dependence of the specific heat on the
temperature in the interval $[0, \infty]$. 
In order to carry out a first check of the curves in picture
\ref{fig_9}, we proceed in a similar way as before when testing the
antiferromagnetic curves. Fig. (\ref{fig_10}) shows the relative
percental differences of the Pad\'e's approximants and
the respective $\beta$-expansion of $C_s (T)$  to the average
Pad\'e for $S=1, 2, 3$ and $5$. We see that the high temperature
series describe well the specific heat of the ferromagnetic
chains for $kT/J \gtrsim 1$.
From theses plots we verify that
the extension of the high-temperature results
to the region of lower temperatures for the 
ferromagnetic case  is worse  than
that of the antiferromagnetic case,
even though integer-spin ferromagnetic materials
are gapless and the behavior of their specific heat about $T=0$ is well
described by the spin-wave approach.

It is also possible to write the magnetic susceptibility per site as a
function of the ground state energy per site $e$. Its low temperature
behavior can be inferred from the extrapolation in a similar fashion
as that of the specific heat per site. Unfortunately, convergence is
not as much as satisfactory. By applying the method of Ref.
\cite{bernu} to the magnetic susceptibility, a large number of
singularities are found in the Pad\'e approximants to this
thermodynamic function. For this reason we use the Dlog-Pad\'e
approximant\cite{buhler} to extend the high temperature expansion of
the antiferromagnetic susceptibility, incorporating the
low-temperature information in a simple way. We follow Ref.
\cite{buhler} to calculate the Dlog-Pad\'e approximant of the magnetic
susceptibility per site $\chi(\beta)$; we refer the reader to this
article for further details on the method. From Ref. \cite{buhler}, we
verify that the Dlog-Pad\'e approximants are independent of the
leading term coefficient of the magnetic susceptibility at low
temperatures, if it vanishes at $T\rightarrow 0$. On the other hand,
if it is non-null  at $T=0$, we need to know the exact coefficient of the
leading term of the magnetic susceptibility for $T\sim 0$ in order to
calculate the Pad\'e approximants at low temperatures. This
also occurs\cite{eggert} for $S= 1/2$.

\begin{figure}[h!t]
\begin{center}
\includegraphics[width=6cm,height=8.5cm,angle=-90]{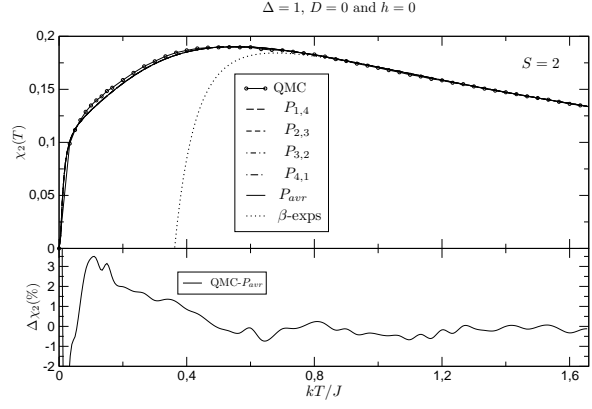}
\caption[fig_11]{The magnetic susceptibility for the antiferromagnetic
case $\Delta=1$, $D=0$ and  $h=0$. a) Dlog-Pad\'e approximants 
of $\chi(T)$ versus $T/J$ are compared to QMC
results\cite{yamamoto96}. b) the average Dlog-Pad\'e approximant
of $\chi(T)$ as a function of $T/J$ for various spin values.}
\label{fig_11}
\end{center}
\end{figure}

Using the QMC method, Yamamoto\cite{yamamoto96} obtained the magnetic
susceptibility of the antiferromagnetic $S=2$ $XXZ$ model ($\Delta =
1$) in the whole interval of temperature of a periodic chain with 96
sites. In Fig. (\ref{fig_11}) we compare the $\beta$-expansion
of $\chi_2 (T)$ and its Dlog-Pad\'e approximants
for $S=2$, with $\Delta=1$, $D=0$ and $h=0$ with Yamamoto's numerical
results. Our extension to lower temperatures assumes that the behavior
of the magnetic susceptibility around $T=0$ be as
$T^{\alpha}\exp(-E_{gap}/kT)$ with $\alpha \approx - 1/4$; the latter value
yields the best fit to the QMC results\cite{yamamoto96}. The bottom of
Fig. (\ref{fig_11}) shows the percental difference between the QMC and
the average Dlog-Pad\'e approximant to the magnetic susceptibility for
$S=2$; it can be seen that $\Delta \chi_2 (\%) \leq 2\%$ up to $kT/J
\thickapprox 0.2$.  From Fig. \ref{fig_11} (top) we verify that our
high-temperature expansion of $\chi_2(T)$ fits well
the numerical result only up to $kT/J \sim 0.7$.

 \begin{figure}[h!t]
\begin{center}
\includegraphics[width=6cm,height=8.5cm,angle=-90]{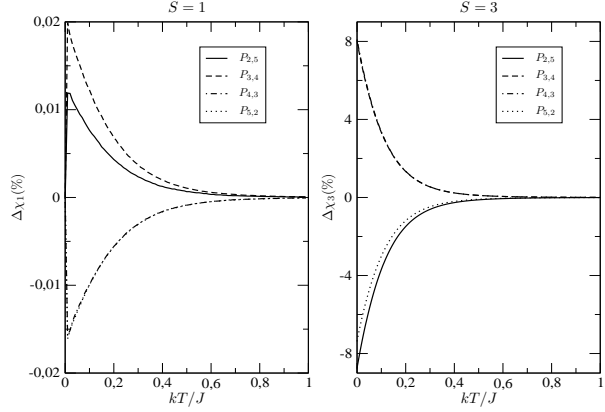}
\caption[fig_12]{The percental difference between the magnetic susceptibility
of the D-log Pad\'e approximants and $P_{avr}$ for the 
antiferromagnetic model ($D=0$), with $S=1$ and $S=3$, and no 
external magnetic fields.}
\label{fig_12}
\end{center}
\end{figure}

Fig. (\ref{fig_12}) shows the deviation from the average $P_{avr}$ of
the Dlog-Pad\'e approximants to the magnetic susceptibility, for the
antiferromagnetic Heisenberg model ($\Delta =1$ and $D=0$) for for
$S=1$ and $S=3$, showing the enhanced extention of our
high-temperature results to lower temperatures. From Ref.
\cite{jolicoeur} we obtain the behavior of $\chi$ around $T=0$
($\alpha = 0.5$); for $S=3$ we assume $ \alpha \thickapprox 0$.

The behavior of the magnetic susceptibility at low temperature for the
ferromagnetic Heisenberg model ($\Delta = -1$ and $D=0$) was obtained
by Takahashi\cite{takahashi} using the modified spin wave theory for
arbitrary spin-S. The magnetic susceptibility for the unitary spin
chain for low temperatures, in units of $k$, is

\begin{align}\label{chi_low}
\chi_s (T)=2\big(\tfrac{S}{S+1}\big)^2T^{-2}\Big(\frac{1}{3}-aT^{1/2}
                 +a^2T+\mathcal O(T^{3/2})\Big),
\end{align}

\noindent where
$a=\frac{\zeta(\tfrac{1}{2})}{2S}\sqrt{\frac{S+1}{\pi}}$ and
$\zeta(x)$ is the Riemann zeta function.
When the Dlog-Pad\'e approximant method\cite{buhler} is applied,
not even the coefficient of the leading term of eq.(\ref{chi_low}) is used,
but only its power. Therefore, a better approximation will be achieved if we
take into account all the three terms in (\ref{chi_low}); 
we do so by applying the two-point Pad\'e 
approximant\cite{baker} method.

We verify from expansion (\ref{chi_low})  that $\chi_s (T)$ is singular at $T=0$. 
Expansion (\ref{free_e}) gives the $\beta$-expansion of $\chi_s (\beta)$
up to order $\beta^6$. The first  terms of  $\chi_s (\beta)$
for   $\Delta = -1$, $D=0$ and $h=0$ are
$\chi_s (\beta) \thickapprox \beta/3+2\beta^2/9$. We define 
the  auxiliary function 
${\mathcal P}_s (\beta) \equiv \frac{\chi_s (\beta)}{\beta^2}-\frac{1}{3\beta}$.
This auxiliary function is regular at $T=0$ and $T\rightarrow \infty$ ($\beta= 0$);
its only problems are that it has non-integer powers 
of $\beta$ and it does not have a Taylor expansion about $T=0$. 
In order to circumvent  this drawback, we proceed as in Ref. \cite{buhler}
and apply the transformation 

\begin{align}
u=\beta^{1/2}/(1+\beta^{1/2})\Leftrightarrow \beta=u^2/(1-u)^2,
          \label{transform}
\end{align}

\noindent where the parameter $\beta\in[0,\infty\rangle$
is mapped  onto $u\in[0,1]$.  
The expansions of ${\mathcal P}$, around $T=0$ $( u=1)$ 
and $\beta=0$ $(u=0)$, are  polynomials in $u$ 
and  therefore we can write its
two-point Pad\'e's approximants, connecting the
expansions of ${\mathcal P}_s$ at $u=1$  and $u=0$.

\begin{figure}[h!t]
\begin{center}
\includegraphics[width=6cm,height=8.5cm,angle=-90]{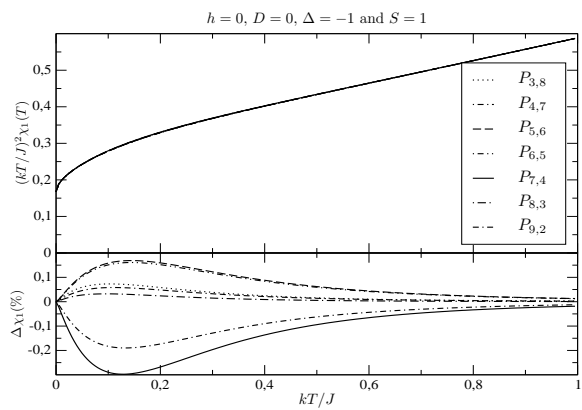}
\caption[fig_13]{(Top) The function $(kT/J)^2 \chi_1 (T)$
versus $kT/J$  of the isotropic ferromagnetic $S=1$ model 
($\Delta = -1$ and $D=0$)  in the absence of external magnetic fields,
for  $P_{avr}$. (Bottom) The relative percental difference 
of non-singular Pad\'e approximants to $P_{avr}$.}
\label{fig_13}
\end{center}
\end{figure}

Fig. (\ref{fig_13}) shows the average Pad\'e of the magnetic
susceptibility for the isotropic ferromagnetic $S=1$ model, in the
absence of an external magnetic field, obtained from the two-point
Pad\'e method and using the transformation (\ref{transform}). For all
Pad\'e approximants, the relative difference to $P_{avr}$ is smaller
than $0.3\%$ in the whole interval of temperature. 

\begin{figure}[h!t]
\begin{center}
\includegraphics[width=6cm,height=8.5cm,angle=-90]{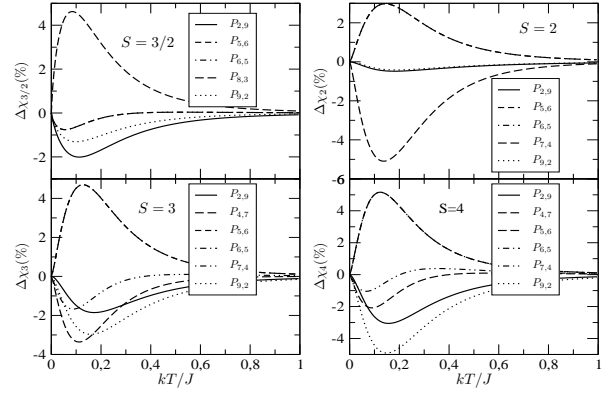}
\caption[fig_14]{The relative percental difference between the
Pad\'e approximants to the magnetic susceptibility
(ferromagnetic case) and $P_{avr}$ for $S= 3/2, 2, 3$
and $4$. Here, $\Delta=-1$, $D=0$ and $h=0$.}
\label{fig_14}
\end{center}
\end{figure}

Finally, Fig. (\ref{fig_14}) shows the relative percental difference
of various Pad\'e approximants to the $P_{avr}$, in the case of the
ferromagnetic isotropic chain model (with $D=0$) and in the absence of
an external magnetic field. The difference increases with $S$;
however, for $S=5$ the relative difference is smaller than $2\%$ up to
$kT/J \thickapprox 0.3$.

\section{Conclusions}

In this report we present the high temperature expansion of the
Helmholtz free energy of the $XXZ$ chain, in the thermodynamic limit,
for an arbitrary spin $S$ up to order $\beta^6$. This expansion is
analytic in $S$ and in $\beta$ and each coefficient of $\beta^n$ ($n=
-1, 0, 1, \cdots 6$) is exact. The model includes anisotropy in the
$z$-direction, single-ion anisotropy and an external magnetic field.
In order to obtain the expansion of the Helmholtz free energy
(\ref{free_e}) we apply the method developed in Ref. \cite{chain_m};
from this expansion we are able to obtain the thermodynamic quantities
of the model for arbitrary spin values. The expansion (\ref{free_e})
recovers the $\beta$-expansion of the Helmholtz free energy presented
in Ref. \cite{jpcm03} for several values of spin, and it is an
extension to the results of Ref. \cite{fukushima}, where we include
the anisotropy effects.

The series in $S$ allows us to obtain the classical limit of the $XXZ$
model, but taking into account the quantum nature of the spins. For
the particular case $h=0$ and $D=0$ we recover Joyce's result for the
classical limit of this model\cite{joyce_prl}. We explicitly show in
expansion \eqref{w_clss} that only the presence of an external
magnetic field distinguishes the classical ferromagnetic and
antiferromagnetic phases in the regime of high temperatures.

We interpolate the $\beta$-expansion with its low temperature series
using the Pad\'e approximant of \eqref{free_e} to extrapolate the
properties of the specific heat to finite temperatures using the
method developed in Ref. \cite{bernu}. Only a few terms of the high
temperature series are needed to yield good agreement: for low spin
values the deviation of all non-singular Pad\'e approximants respect
to $P_{avr}$ fluctuates around 5\%. 

For the antiferromagnetic magnetic susceptibility we use the
Dlog-Pad\'e\cite{buhler} approximants to investigate the finite
temperature properties of the spin-$s$ quantum XXZ chain. In order to
plot the magnetic susceptibility $\chi_s (T)$ as a function of
temperature for different spin values we take the average of all
non-singular Dlog-Pad\'e approximants, since their deviation from the
corresponding $P_{avr}$ is less than 2\%. The ferromagnetic case is
studied by applying two-point like Pad\'e approximants, since we have
the first three terms of the low temperature series
expansion\cite{takahashi}. The magnetic susceptibility is plotted for
different spins, in the same way as for the previous case.

Obviously, if we knew more low temperature information about any
physical quantities, it could be also interpolated with our high
temperature expansions, by adapting different approximants available
in the literature. Our main limitation regarding the study of finite
temperature behavior of other parameters of the Hamiltonian
\eqref{hamiltonian} is the lack of related low temperature
information.

\begin{acknowledgments}

O. R. thanks FAPEMIG for financial support.
S.M. de S. and O.R. thank FAPEMIG and CNPq for partial financial support.
M.T.T. thank    s CNPq and FAPERJ for partial financial
support.

\end{acknowledgments}

\appendix
\section{normalized traces of  $(S^z)^l$}\label{appendixA} 

We have that $\mathrm{tr}((S^z)^n) = 0$ if $n$ 
is odd.  For the purposes of this appendix, $n$ is taken as an even integer.
In order to calculate  $\mathrm{tr}((S^z)^n)$ we use the result of  
Ref.\cite{abramowitz} that  permits us to  write this trace  as a function 
of the spin $S$,

\begin{align}\label{sn_nat}
\mathrm{tr}((S^z)^n)=\sum_{j=-S}^{S}j^n=\sum_{r=0}^{n+1}C_{r,n}S^r,
\end{align}

\noindent where $S= 1/2, 1, 3/2, \cdots$, 
$C_{r,n}=\frac{(1+ (-1)^n)}{n+1}\sum_{k=0}^{n+1}\binom{k}{r}b_{k, n+1}$,
so that $b_{k, n}$'s are the coefficients of  Bernoulli's 
polynomial $B_n(x)$. They are  defined by\cite{abramowitz}

\begin{align}
B_{n}(x) = \sum_{k=0}^{n} b_{k,n} x^k.
\end{align}

On the other hand,  we prefer to rewrite  eq.\eqref{sn_nat}
as a function of the square of the norm of the spin ${\bf S}$ 
(that is, $S(S+1)$),  which is a constant of motion,

\begin{align}\label{sn_fac}
\sum_{j=-S}^{S}j^n=(2S+1)\sum_{r=0}^{n/2}A_{r,n}(S(S+1))^r.
\end{align}

\noindent We remind that $2S+1$ is equal to the dimension
of the Hilbert space at each site, that is,
${\rm tr}_i({\bf 1}) = 2S+1$, $i= 1, 2, \cdots, N$. Comparing 
eqs.\eqref{sn_nat} and \eqref{sn_fac},
and after some algebraic manipulation,  we
obtain each coefficient $A_{r,n}$ of  eq.\eqref{sn_fac}
as a combination of the Bernoulli numbers\cite{abramowitz} $B_{n+1-j}$,

\begin{align} \label{A3}
A_{k,n}=\frac{(1+ (-1)^n)(-1)^{k+1}}{n+1}\sum_{j=0}^{k}\binom{n+1}{j}
\binom{2k-j}{k} B_{n+1-j}.
\end{align}

In eq.(\ref{hamiltonian}) we rewrite Hamiltonian (\ref{hamt1})
in terms of the rescaled spin operator ${\bf s}$. To obtain the results
(\ref{B2})-(\ref{B8}) we need to calculate the powers of the operator
$s^z$. From the results \eqref{sn_nat}-\eqref{A3} we are able to
 write the normalized traces of $s^z$, that is
$\langle (s^z)^{2n}\rangle$ just as a polynomial of 
$y\equiv \frac{1}{S(S+1)}$ of degree $n/2-1$, which 
 reads

\begin{align}
\langle(s^z)^{n}\rangle=  \sum_{r=0}^{n/2}A_{n/2 - r, n} \; y^r.
\end{align}

In terms of the  normalized matrix 
representation, the commutation relation  becomes

\begin{align} 
 [s^x,s^y]=\frac{i s^z}{\sqrt{S(S+1)}} = i \sqrt{y} \ s^z ,
\end{align}

\noindent which is useful  to evaluate the normalized traces.

\section{The $\beta$-expansion of the free energy for the  
spin-$s$ $XXZ$ chain} \label{B}

We call ${\mathcal W}_s (\beta)$ the Helmholtz free energy of 
the Hamiltonian  (\ref{hamiltonian}), that drives the dynamics
 of the quantum spin chain with  
unitary norm. Let ${\mathcal W}_S (\beta)$ be the Helmholtz 
free energy  of the Hamiltonian

\begin{align}\label{B1}
{\bf H}_S  = \sum_{i=1}^N J \left( {\bf S}_{i}, {\bf S}_{i+1}\right)_{\Delta} 
       - h S^{z}_{i} + D (S^{z}_{i})^2,    
\end{align}

\noindent where the norm of the spin-${\bf S}$ is $\sqrt{S(S+1)}$ and $S= 1/2, 1, 3/2, \cdots$. 
In Ref. \cite{jpcm03} we 
calculated the high temperature expansion
of the Helmholtz free energy of the Hamiltonian (\ref{B1}) up to
$S=4$.  The relation between the free energies derived from Hamiltonians 
(\ref{hamiltonian}) and (\ref{B1}) is  

\begin{subequations} \label{b2}
\begin{eqnarray}
{\mathcal W}_s ( J,  \Delta, h, D; \beta) & = & 
\frac{1}{S(S+1)}  {\mathcal W}_S \left( J, \Delta, \sqrt{S(S+1)}  h,D; 
\frac{ \beta}{S(S+1)}\right)
                                          \label{b2a}  \\
%
%
& = & {\mathcal W}_S \left( \frac{J}{S(S+1)},  \Delta, \frac{h}{\sqrt{S(S+1)}}, 
  \frac{D}{S(S+1)};  \beta \right).
                                          \label{b2b}
\end{eqnarray}
\end{subequations}

\noindent In eqs. (\ref{b2}),  the parameters of the respective
 Hamiltonians are shown explicitly.  Equating eqs. (\ref{b2a}) 
 and (\ref{b2b}), we verify that ${\mathcal W}_S$ 
 is an homogeneous function of degree 1.

In this appendix we present the high temperature expansion
of the Helmholtz free energy ${\mathcal W}_s (\beta)$,
up to order $\beta^6 \, (n=7)$, for arbitrary value of $s (s=S)$. This expansion
can be written in terms of  $y\equiv \frac{1}{S(S+1)}$, that is,

\begin{align}\label{free_e}
{\mathcal W}_s (\beta)=-\frac{\ln(2S+1)}{\beta}+\sum_{r=0}^{6}w_{r}(y)\beta^r
+\mathcal{O}(\beta^7).
\end{align}

\noindent The coefficients  $w_{r}(y)$, for $r= 0, 1,\cdots, 6$,
are  shown below.  For the sake of simplicity, we define
$x \equiv (3y-4)$.

\begin{align}\label{B2}
w_{0}(y)=\tfrac{D}{3}
\end{align}
\begin{align}
w_{1}(y)= \tfrac {D^{2}\,x}{90}  - 
{\displaystyle \tfrac {h^{2}}{6}}  - {\displaystyle \tfrac {J^{2}}{
9}}  - {\displaystyle \tfrac {J^{2}\,\Delta ^{2}}{18}} 
\end{align}

\begin{align}
w_{2}(y)= &({\displaystyle \tfrac {J^{2}\,{D}}{135}}  + 
{\displaystyle \tfrac {{D}^{3}\,(15\,y - 4)}{5670}}  -
{\displaystyle \tfrac {J^{2}\,\Delta ^{2}\,{D}}{135}}  -
{\displaystyle \tfrac {h^{2}\,{D}}{90}} )\,x -
{\displaystyle \tfrac {J^{3}\,\Delta \,y}{36}}  + {\displaystyle
\tfrac {J\,\Delta \,h^{2}}{9}}
\end{align}

\begin{align}
w_{3}(y)=&({\displaystyle \tfrac {2\,J\,\Delta \,h^{2}\,{D}}{135}}
 - {\displaystyle \tfrac {h^{2}\,{D}^{2}\,(15\,y - 4)}{3780
}}  - {\displaystyle \tfrac {J^{2}\,{D}^{2}\,(27\,y + 4)}{
18900}}  + {\displaystyle \tfrac {{D}^{4}\,(105\,y^{2} - 54
\,y - 8)}{113400}}  - {\displaystyle \tfrac {8\,J^{2}\,\Delta ^{2}\,{D}
^{2}\,(2\,y - 1)}{4725}} )x\mbox{} \nonumber\\
& + {\displaystyle \tfrac {h^{4}
\,(y + 2)}{360}}  - {\displaystyle \tfrac {J^{4}\,( - 128\,y + 12
 + 33\,y^{2})}{16200}}  + {\displaystyle \tfrac {J^{2}\,\Delta ^{2}\,h^{2}\,(3\,y
 - 14)}{270}}  - {\displaystyle \tfrac {J^{4}\,\Delta ^{4}\,( - 16
\,y + 14 + y^{2})}{5400}} \nonumber\\
&  + {\displaystyle \tfrac {J^{2}\,h^{2}\,
(9\,y + 8)}{540}}  - {\displaystyle \tfrac {J^{4}\,\Delta ^{2}\,( - 32\,y -
72 + 27\,y^{2})}{8100}}
\end{align}

\begin{align}
w_{4}(y)= &({\displaystyle \tfrac {J^{2}\,\Delta ^{2}\,h^{2}\,{D}\,(16
\,y - 43)}{4725}}  + {\displaystyle \tfrac {16\,J\,\Delta \,h^{2}
\,{D}^{2}\,(2\,y - 1)}{4725}}  - {\displaystyle \tfrac {J^{
4}\,\Delta ^{4}\,{D}\,(y^{2} - 12\,y + 4)}{11340}}  
- {\displaystyle \tfrac {h^{2}\,{D}^{3}\,(105\,y^{2} - 54\,y - 8)}{56700}} 
                                  \nonumber\\
& - {\displaystyle \tfrac {J^{3}\,\Delta \,\mathrm{D
}^{2}\,y\,(9\,y - 2)}{8100}} + {\displaystyle \tfrac {{D}^{5}\,(
15\,y - 4)\,(105\,y^{2} - 48\,y - 16)}{3742200}}   - {\displaystyle \tfrac {J^{4}\,
\Delta ^{2}\,{D}\,(3\,y + 4)\,(y - 4)}{22680}}
+ {\displaystyle \tfrac {J^{2}\,h^{2}\,{D}\,(27\,y
 + 4)}{18900}}  \nonumber\\
&   +
{\displaystyle \tfrac {J^{4}\,{D}\,(5\,y^{2} - 32\,y - 8)}{
22680}}   + {\displaystyle \tfrac {J^{2}\,{D}^{3}\,(3\,y^{2}
 - 60\,y - 32)}{170100}}  + {\displaystyle \tfrac {h^{4}\,\mathrm{
D}\,(5\,y + 8)}{7560}}  - {\displaystyle \tfrac {J^{2}\,\Delta ^{2}\,{D}^{3
}\,(75\,y^{2} - 63\,y + 4)}{42525}} )x \nonumber\\
& - {\displaystyle \tfrac {J\,\Delta \,h^{4}\,(y + 2)}{135}
}  + {\displaystyle \tfrac {J^{3}\,\Delta \,h^{2}\,(9\,y + 8)\,(y
 - 3)}{1350}}  + {\displaystyle \tfrac {J^{3}\,\Delta ^{3}\,h^{2}
\,( - 48\,y + 92 + 3\,y^{2})}{4050}} - {\displaystyle \tfrac {J^{5}\,\Delta ^{3}\,y\,( - 16\,y
 - 4 + 3\,y^{2})}{6480}}  \nonumber\\
&  - {\displaystyle \tfrac {J^{5}\,\Delta
\,y\,( - 16\,y - 4 + 3\,y^{2})}{3240}}
\end{align}

\begin{align}
w_{5}(y)=& ( - {\displaystyle \tfrac {J^{2}\,{D}^{4}\,(65223\,y^{3} +
34020\,y^{2} - 5904\,y - 16064)}{785862000}}  + {\displaystyle
\tfrac {J^{3}\,\Delta ^{3}\,h^{2}\,{D}\,(3\,y^{2} - 36\,y
 + 40)}{8505}}  + {\displaystyle \tfrac {
J^{5}\,\Delta \,{D}\,y^{2}\,(4\,y - 27)}{40500}}   \nonumber\\
& - {\displaystyle \tfrac {J^{4}\,{D}^{2}\,(4287\,y^{
3} - 14910\,y^{2} + 2404\,y + 3184)}{23814000}}  +
{\displaystyle \tfrac {J^{3}\,\Delta \,h^{2}\,{D}\,(39\,y^{
2} - 62\,y - 40)}{28350}} - {\displaystyle \tfrac {2\,J\,\Delta \,h^{4}\,{D}
\,(16\,y + 27)}{14175}}   \nonumber\\
& + {\displaystyle \tfrac {2\,J\,\Delta \,h^{2}\,{D}
^{3}\,(75\,y^{2} - 63\,y + 4)}{42525}}  - {\displaystyle \tfrac {J
^{3}\,\Delta \,{D}^{3}\,y\,(213\,y^{2} - 129\,y + 20)}{
425250}} + {\displaystyle \tfrac {J^{2}\,h^{2}\,{D}^{2}\,(
123\,y^{2} + 32\,y + 32)}{113400}} \nonumber\\
& + {\displaystyle \tfrac {{D}^{6}\,(7182945\,y^{4}
 - 6404130\,y^{3} + 569016\,y^{2} + 443424\,y + 2944)}{
30648618000}}  - {\displaystyle \tfrac {h^{2}\,{D}^{4}\,(15\,y - 4
)\,(105\,y^{2} - 48\,y - 16)}{1496880}} \nonumber\\
& - {\displaystyle \tfrac {J^{4}\,\Delta ^{4}\,{D}^{2
}\,(111\,y^{3} - 1155\,y^{2} + 888\,y - 16)}{1190700}}  +
{\displaystyle \tfrac {h^{4}\,{D}^{2}\,(105\,y^{2} + 74\,y
 - 72)}{226800}} + {\displaystyle \tfrac {J^{2}
\,\Delta ^{2}\,h^{2}\,{D}^{2}\,(15\,y^{2} - 36\,y + 16)}{
5670}} \nonumber\\
& - {\displaystyle \tfrac {J^{4}\,\Delta ^{2}\,{D}^{2
}\,(369\,y^{3} - 546\,y^{2} - 751\,y - 116)}{1488375}}  - {\displaystyle \tfrac {J^{2}\,\Delta
^{2}\,{D}^{4
}\,(204885\,y^{3} - 224028\,y^{2} + 39888\,y + 8768)}{196465500}
}\nonumber\\
& -
{\displaystyle \tfrac {J^{5}\,\Delta ^{3}\,{D}\,y^{2}\,(4\,
y - 27)}{40500}}    )x
  + {\displaystyle \tfrac {J^{4}\,
\Delta ^{4}\,h^{2}\,( - 768\,y^{2} - 2872 + 2748\,y + 45\,y^{3})
}{340200}} - {\displaystyle \tfrac {h^{6
}\,(12\,y + 16 + 3\,y^{2})}{45360}}   \nonumber\\
&  -
{\displaystyle \tfrac {J^{2}\,h^{4}\,(632\,y + 432 + 333\,y^{2})}{
226800}}  - {\displaystyle \tfrac {J^{6}\,\Delta ^{2}\,(101120 +
203200\,y + 83481\,y^{2} - 289584\,y^{3} + 42174\,y^{4})}{
142884000}}  \nonumber\\
& - {\displaystyle \tfrac {J^{6}\,( - 33760 - 1920\,y +
239418\,y^{2} - 127872\,y^{3} + 14067\,y^{4})}{214326000}}  - {\displaystyle \tfrac
{J^{2}\,\Delta ^{2}\,h^{4}\,( -
424\,y - 1074 + 69\,y^{2})}{56700}}    \nonumber\\
&  - {\displaystyle \tfrac {J^{6}\,\Delta ^{4}\,( - 16960 +
50720\,y + 30846\,y^{2} - 33624\,y^{3} + 4239\,y^{4})}{71442000}
} + {\displaystyle \tfrac {J^{4}\,h^{2}\,( - 108\,y^{2} - 32
 - 160\,y + 27\,y^{3})}{45360}}    \nonumber\\
& - {\displaystyle \tfrac {J^{6}\,\Delta ^{6}\,(1712 - 15408
\,y + 14688\,y^{2} - 2700\,y^{3} + 135\,y^{4})}{42865200}} 
+ {\displaystyle \tfrac {J^{4}\,\Delta^{2}\,h^{2}\,
( -813\,y^{2} + 848 + 383\,y + 135\,y^{3})}{85050}}
\end{align}

\begin{align}\label{B8}
w_{6}(y)= & - {\displaystyle \tfrac {2\,J\,\Delta \,h^{4}\,{D}^{2}\,x
^{2}}{4725}}  + ( - {\displaystyle \tfrac {J^{2}\,\Delta ^{2}\,
{D}^{5}\,(1780695\,y^{4} - 2280645\,y^{3} + 677964\,y^{2}
 + 89296\,y - 25024)}{2554051500}}  \nonumber\\
& + {\displaystyle \tfrac {J^{2}\,h^{2}\,{D}^{3}\,(
262035\,y^{3} - 85176\,y^{2} + 12576\,y - 16064)}{392931000}}
- {\displaystyle \tfrac {J^{6}\,\Delta ^{6}\,{D}\,(
45\,y^{4} - 720\,y^{3} + 2988\,y^{2} - 1608\,y - 208)}{15309000}
}  \nonumber\\
& - {\displaystyle \tfrac {J^{2}\,h^{4}\,{D}\,(31\,y^{2}
 + 40\,y + 16)}{75600}}  - {\displaystyle \tfrac {J^{4}\,\Delta ^{2}\,{D}^{3
}\,(25002\,y^{4} - 47646\,y^{3} - 23907\,y^{2} + 15308\,y + 10464
)}{147349125}}  \nonumber\\
& - {\displaystyle \tfrac {J^{6}\,\Delta ^{4}\,{D}\,(
729\,y^{4} - 5328\,y^{3} + 4488\,y^{2} + 7784\,y + 1248)}{
30618000}}  + {\displaystyle \tfrac {J^{4}\,h^{2}\,{D}\,(1935\,
y^{3} - 1974\,y^{2} + 2404\,y + 3184)}{23814000}}  \nonumber\\
& + {\displaystyle \tfrac {J^{6}\,\Delta ^{2}\,{D}\,(
108\,y^{4} - 684\,y^{3} + 201\,y^{2} + 524\,y + 624)}{15309000}}
 + {\displaystyle \tfrac {J^{2}\,\Delta ^{2}\,h^{2}\,
{D}^{3}\,(40977\,y^{3} - 91791\,y^{2} + 58890\,y - 7240)}{
19646550}}  \nonumber\\
& + {\displaystyle \tfrac {J^{3}\,\Delta \,h^{2}\,{D}
^{2}\,(7425\,y^{3} - 11907\,y^{2} - 2928\,y - 928)}{5953500}}
+ {\displaystyle \tfrac {J^{4}\,\Delta ^{2}\,h^{2}\,
{D}\,(972\,y^{3} - 4851\,y^{2} + 1567\,y + 2084)}{1786050}
}  \nonumber\\
& - {\displaystyle \tfrac {J^{2}\,\Delta ^{2}\,h^{4}\,
{D}\,(35\,y^{2} - 141\,y - 332)}{56700}}  
+ {\displaystyle \tfrac {J\,\Delta \,h^{2}\,{D}^{4}
\,(204885\,y^{3} - 224028\,y^{2} + 39888\,y + 8768)}{98232750}}
 \nonumber\\
& - {\displaystyle \tfrac {J^{5}\,\Delta ^{3}\,{D}^{2
}\,y\,(1422\,y^{3} - 8253\,y^{2} + 3697\,y - 208)}{11907000}}
 - {\displaystyle \tfrac {J^{5}\,\Delta \,{D}^{2}\,y
\,(1455\,y^{3} - 5418\,y^{2} - 428\,y + 376)}{11907000}}  \nonumber\\
& + {\displaystyle \tfrac {J^{4}\,\Delta ^{4}\,h^{2}\,
{D}\,(1665\,y^{3} - 22113\,y^{2} + 53808\,y - 35212)}{
17860500}}  + {\displaystyle \tfrac {h^{4}\,{D}^{3}\,(2625\,y^{
3}+ 300\,y^{2} - 1928\,y + 224)}{7484400}} \nonumber\\
&  + {\displaystyle \tfrac {J^{3}\,\Delta ^{3}\,h^{2}\,
{D}^{2}\,(333\,y^{3} - 3465\,y^{2} + 4596\,y - 1504)}{
893025}}   + {\displaystyle \tfrac {J^{6}\,{D}\,(603\,y^{4} -
5400\,y^{3} + 10062\,y^{2} + 3520\,y - 416)}{30618000}}  \nonumber\\
& - {\displaystyle \tfrac {J^{4}\,\Delta ^{4}\,{D}^{3
}\,(2\,y - 1)\,(8925\,y^{3} - 79200\,y^{2} + 49076\,y + 8912)}{
196465500}} - {\displaystyle \tfrac {J^{3}\,\Delta \,{D}^{4}\,y
\,(4413\,y^{3} - 4214\,y^{2} + 724\,y + 48)}{11907000}}   \nonumber\\
&
 - {\displaystyle \tfrac {J^{2}\,{D}^{5}\,(150639\,y
^{4} + 58088\,y^{3} - 28064\,y^{2} - 67200\,y - 17664)}{
3405402000}} - {\displaystyle
\tfrac {h^{6}\,{D}\,(7\,y^{2} + 22\,y + 24)}{226800}}  \nonumber\\
& - {\displaystyle \tfrac {J^{4}\,{D}^{3}\,(6201\,y^{
4} + 346320\,y^{3} - 363672\,y^{2} - 223664\,y - 35136)}{
2357586000}}  - {\displaystyle \tfrac {J\,\Delta \,h^{4}\,{D}
^{2}\,y^{2}}{567}}  \nonumber\\
& + {\displaystyle \tfrac {{D}^{7}\,(15\,y - 4)\,(
315315\,y^{4} - 238140\,y^{3} + 1896\,y^{2} + 17664\,y + 2944)}{
30648618000}}  \nonumber\\
& - {\displaystyle \tfrac {h^{2}\,{D}^{5}\,(7182945\,
y^{4} - 6404130\,y^{3} + 569016\,y^{2} + 443424\,y + 2944)}{
10216206000}}  )x  + {\displaystyle \tfrac {J\,\Delta \,h^{6}\,(44\,y + 54 +
11\,y^{2})}{28350}}
\nonumber\\
& - {\displaystyle \tfrac {J^{7}\,\Delta \,y\,(1600 + 6432\,
y + 51631\,y^{2} - 24768\,y^{3} + 2556\,y^{4})}{40824000}}
- {\displaystyle \tfrac {J^{3}\,\Delta\,h^{4}\
,( - 284\,y^{2} - 640 - 834\,y + 99\,y^{3})}{113400}}   \nonumber\\
& - {\displaystyle \tfrac {J^{7}\,\Delta ^{5}\,y\,( - 400 -
3344\,y + 9228\,y^{2} - 2964\,y^{3} + 243\,y^{4})}{20412000}}
 - {\displaystyle \tfrac {J^{3}\,\Delta ^{3}\,
h^{4}\,( - 228\,y^{2} + 1488 + 328\,y + 15\,y^{3})}{85050}} \nonumber\\
& + {\displaystyle \tfrac {J^{5}\,\Delta ^{3}\,h^{2}\,( -
33440 + 1080\,y + 51411\,y^{2} - 25974\,y^{3} + 2754\,y^{4})}{
7144200}}    + {\displaystyle \tfrac {J^{5}\,\Delta \,h^{2}\,(10112 +
30512\,y + 4629\,y^{2} - 20844\,y^{3} + 3159\,y^{4})}{7144200}}
 \nonumber\\
&+ {\displaystyle \tfrac {J^{5}\,\Delta ^{5}\,h^{2}\,(20528
 - 31872\,y + 16452\,y^{2} - 2700\,y^{3} + 135\,y^{4})}{7144200}
} - {\displaystyle \tfrac {J^{7}\,\Delta ^{3}\,y\,(3200 +
19808\,y + 14719\,y^{2} - 12912\,y^{3} + 1584\,y^{4})}{20412000}
}
\end{align}

The coefficients $J$, $D$ and $h$ in (\ref{B2})-
(\ref{B8}) are the constants in Hamiltonian (\ref{hamiltonian}).

Writing the expansion (\ref{free_e}) in terms of the variable $y$
greatly simplifies the calculation of its classical limit
($y\rightarrow 0$).

\end{document}